\documentclass[prd,aps,12pt]{revtex4}
\usepackage{graphicx}
\usepackage{latexsym}
\usepackage{amsmath}
\begin{document}

\title{Quantum magnetism and criticality}

\author{Subir Sachdev}
\affiliation{Department of Physics, Harvard University, Cambridge
MA 02138}
\date{\today\\[24pt]}

\begin{abstract}
Magnetic insulators have proved to be fertile ground for studying new types of 
quantum many body states, and I survey recent experimental and theoretical examples.
The insights and methods transfer also to novel superconducting and metallic states.
Of particular interest are critical quantum states, sometimes found at quantum phase transitions, 
which have gapless excitations with no particle-
or wave-like interpretation, and control a significant portion of the finite temperature
phase diagram. Remarkably, their theory is connected to holographic descriptions
of Hawking radiation from black holes.
\end{abstract}

\maketitle

\section{Introduction}
\label{sec:intro}

Quantum mechanics introduced several revolutionary ideas into physics, but perhaps the
most surprising to our classical intuition was that of linear superposition. This is the 
idea that any linear combination of two allowed quantum states of a system is also
an allowed state. For single particles, linear superposition allows us to explain experiments like
double-slit interference, in which electrons can choose one of two possible slits in an obstacle, but instead
choose a superposition of the two paths, leading to an interference pattern in their probability to
traverse the obstacle. However the concept of linear
superposition had unexpected consequences when extended 
to many particles. In a well-known paper, Einstein, Podolsky, and Rosen \cite{epr} (EPR) 
pointed out that even with just two electrons, linear superposition implied non-local and non-intuitive features of quantum theory. They imagined a state in which the spin of one electron 
was up and the second down, which we write as $\left|\uparrow \downarrow \right\rangle$. Now let us take
a linear superposition of this state with one with the opposite orientations to both electron spins, leading to the state $(\left|\uparrow \downarrow \right\rangle - \left|\downarrow \uparrow \right\rangle)/\sqrt{2}$ Such a `valence bond' state is similar to that found in many situations in chemistry and solid state physics {\em e.g.\/} the state of the two electrons in a hydrogen molecule. EPR imagined a `thought experiment' in which
the two electrons are moved very far apart from each other, while the quantum states of their
spins remains undisturbed. Then we have the remarkable prediction that if measure the spin of 
one of the electrons, and find to be $\left|\uparrow \right\rangle$ (say), 
the perfect anticorrelation of the spins in the valence bond implies that
the spin of the other far-removed electron is immediately fixed to be $\left|\downarrow \right\rangle$. EPR found this sort
of `action-at-a-distance' effect unpalatable. Today, we have many examples of 
the EPR experiments in the laboratory, and the peculiar non-local features of quantum mechanics
have been strikingly confirmed. In recent years, the term `entanglement' has frequently been used
to describe the non-local
quantum correlations inherent in the valence bond state of two spins.

The purpose of this article is to describe the consequences of the linear superposition principle of quantum mechanics
when extended to a very large number of electrons (and other particles). We are interested in how many 
electrons entangle with each other in the low energy states of Hamiltonians similar to those found in certain transition metal
compounds, and related `correlated electron' systems. While there is a precise definition of entanglement for two
electrons which captures the essence of the EPR effect, much remains to be understood in how to define an analogous
measure of `quantum non-locality' for large numbers of electrons. We shall sidestep this issue by restating the question: how
can we tell when two quantum states have genuinely distinct superpositions of the electron configurations ?
Further, can we classify and list the distinct classes of states likely to be found in correlated electron materials ?

A useful strategy is to follow the low energy states of a Hamiltonian which varies as a function  of a parameter $g$. 
We examine whether the ground states breaks a symmetry of the Hamiltonian. We also examine the quantum numbers
of the excitations needed to describe the low energy Hilbert space. If one or more of such `discrete' characteristics is
distinct between large and small $g$, then we assert that these limiting states realize distinct quantum phases.
The distinction between the phases implies, in particular, that they cannot be smoothly connected as $g$ is
varied. Consequently there must be at least one special value of $g$, which we denote $g_c$, at which there is a
non-analytic change in characteristics of the ground state {\em e.g.\/} the ground state energy. Often some of the
most interesting quantum states, with non-trivial and non-local types of entanglement, are found precisely
at the quantum critical point $g=g_c$, where the system is delicately balanced between the states
of either side of $g_c$. This happens when there is a correlation length which becomes large near
$g_c$, as is the case close to a ``second-order'' phase transition (near strong ``first-order'' transitions the ground state
simply jumps discontinuously between simple limiting states on either side of $g_c$). 
Thus a description of the Hamiltonian as a function of $g$ exposes
two quantum phases and their connection across a quantum phase transition, and we present several examples
of such phenomena below. 

Experimentally, it is almost never possible to examine the ground state of a large quantum system in isolation.
There are small random perturbations from the environment which eventually bring the system into equilibrium
at a temperature $T >0$. So it is crucial to examine the fingerprints of the ground state quantum phases and quantum
critical points on the non-zero temperature phase diagram. We will turn to this issue in Section~\ref{sec:qc}, where we
will argue that such fingerprints can often be significant and easily detectable. Indeed, the quantum critical point can
serve as the best point of departure for understanding the entire phase diagram in the $g$, $T$ plane.

We will begin in Section~\ref{sec:ins} by describing the rich variety of quantum phases that appear in
two-dimensional insulators, whose primary excitations are $S=1/2$ electronic spins on the sites of a lattice. 
Interest in such two-dimensional spin systems was initially stimulated by
the discovery of high temperature superconductivity in the cuprate compounds. However, since then, such quantum spin systems
have found experimental applications in a wide variety of systems, as we will describe below. 
The critical quantum phases and points
of such materials will be discussed in Section~\ref{sec:qcp}. Section~\ref{sec:ms} will extend
our discussion to include charge fluctuations, and describe some recent proposals for novel metallic and superconducting states,
and associated quantum critical points. 

Although we will not discuss these issues here, related ideas apply also in other spatial dimensions. Similarly quantum states appear in 
three dimensions, although the quantum critical points often have a distinct character. A fairly complete understanding of one dimensional
correlated states has been achieved, building upon `bosonization' methods. 

\section{Phases of insulating quantum magnets}
\label{sec:ins}

High temperature superconductivity appears when insulators like La$_2$CuO$_4$ are doped with mobile
holes or electrons. La$_2$CuO$_4$ is an antiferromagnet, in which the magnetic degrees of freedom are
$S=1/2$ unpaired electrons, one on each Cu atom. The Cu atoms reside on the vertices of square lattices, which 
are layered atop each other in the three-dimensional crystal. The couplings between the layers are negligible (although this small
coupling is important in obtaining a non-zero magnetic ordering temperature),
and so we need only consider the $S=1/2$ spins ${\bf S}_i$, residing on the sites of a square lattice, $i$. The spins are coupled
to each other via a superexchange interaction, and so we will consider here the Hamiltonian
\begin{equation}
H_0 = \sum_{\langle ij \rangle} J_{ij} {\bf S}_i \cdot {\bf S}_j 
\label{h0}
\end{equation}
Here $\langle ij \rangle$ represents nearest-neighbor pairs, and $J_{ij} >0$ is the exchange interaction.
The antiferromagnetism is a consequence of the positive value of $J_{ij}$ which prefers an anti-parallel alignment
of the spins. We will initially consider, in Section~\ref{sec:neel} 
the case of $i$ on the sites of a square lattice, and all $J_{ij}=J$ equal, but will generalize in the subsequent sections
to other lattices and additional interactions between the spins.

Our strategy here will be to begin by guessing possible phases of $H_0$, and its perturbations, in some simple limiting cases.
We will then describe the low energy excitations of these states by quantum field theories. By extending
the quantum field theories to other parameter regimes, we will see relationships between the phases,
discover new phases and critical points, and generally obtain a global perspective on the phase diagram.

\subsection{N\'eel ordered states}
\label{sec:neel}

With all $J_{ij}=J$, there is no parameter, $g$, which can tune the ground state of $H_0$;
$J$ sets the overall energy scale, but does not modify the wavefunction of any
of the eigenstates. So there is only a single quantum phase to consider.

\begin{figure}[htbp]
  \centering
  \includegraphics[width=3in]{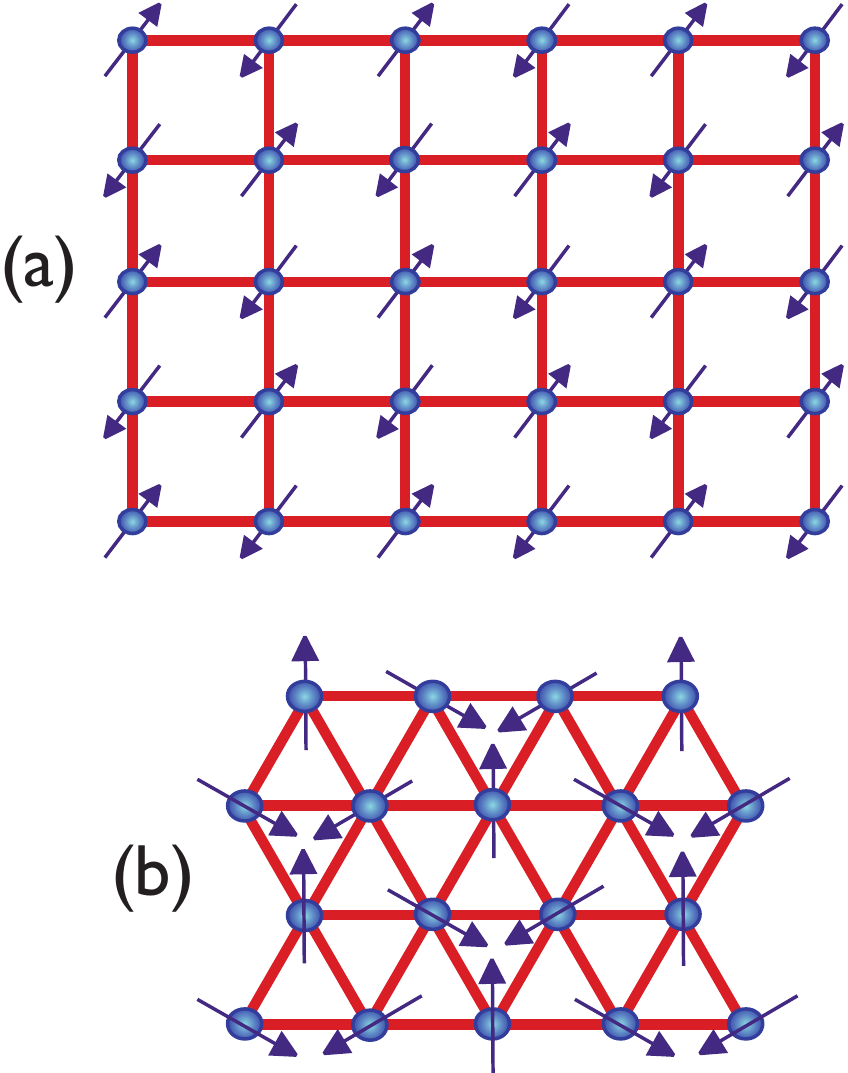}
  \caption{N\'eel ground state of the $S=1/2$ antiferromagnet $H_0$ 
  will all $J_{ij} = J$ on the (a) square and (b) triangular lattices.
  The spin polarization is (a) collinear and (b) coplanar.}
  \label{neel}
\end{figure}

For the square lattice, there is convincing numerical and experimental evidence \cite{matsumoto} that the ground
state of $H_0$ has N\'eel order, as illustrated in Fig.~\ref{neel}a. 
This state spontaneously
breaks the spin rotation symmetry of $H_0$, and is characterized by the expectation value
\begin{equation}
\langle {\bf S}_j \rangle = (-1)^j  \boldsymbol{\Phi}
\label{sPhi}
\end{equation}
Here $(-1)^j$ represents the opposite orientations of the spins on the two sublattices
as shown in Fig~\ref{neel}a. The vector $\boldsymbol{\Phi}$  represents the orientation and magnitude of the 
N\'eel order. We have $|\boldsymbol{\Phi}| < 1/2$, implying that the ground state wavevefunction is not simply
the classical state sketched in Fig.~\ref{neel}a, but has quantum fluctuations about it. These fluctuations will entangle the spins
with each other, but qualitatively the essential character of the state is captured by the pattern in Fig.~\ref{neel}a.
Note that this pattern implies a long-range, and classical, correlation between the spins, but not a significant
amount of entanglement because there are no EPR effects between any pair of well separated spins.

Having described the ground state, we now turn to a consideration of the excitations. This is most conveniently done by
writing down the form of the Feynman path integral for the trajectories of all the spins in imaginary time, $\tau$. After taking
the long-wavelength limit to a continuous two-dimensional space, this path integral defines a quantum field theory
in 2+1 dimensional spacetime with co-ordinates $(r,\tau)$. The quantum field theory is for a field $\boldsymbol{\Phi} (r,\tau)$,
which is the value of the N\'eel order when averaged over the square lattice spins, ${\bf S}_i$ located on sites within
some averaging neighborhood of $r$.
A quick derivation of the effective action for this quantum field theory is provided by 
writing down all terms in powers and gradients of $\boldsymbol{\Phi}$ which are invariant under all the symmetries
of the Hamiltonian. In this manner we obtain the action \cite{chn,csy}
\begin{equation}
\mathcal{S}_{\boldsymbol{\Phi}} = \int d^2 r d \tau \left(  (\partial_\tau \boldsymbol{\Phi})^2
+ v^2 (\nabla_x \boldsymbol{\Phi} )^2 + s \boldsymbol{\Phi}^2  + u (\boldsymbol{\Phi}^2)^2 \right)
\label{SP}
\end{equation}
Here $v$ is a spin-wave velocity, and $s,u$ are parameters whose values adjusted to obtain N\'eel order in the
ground state. In mean-field-theory, this happens for $s<0$, where we have $|\langle \boldsymbol{\Phi} \rangle | = (-s)/(2u)$
by minimization of the action $\mathcal{S}_{\boldsymbol{\Phi}}$. A standard computation of the fluctuations about
this saddle point shows that the low energy excitations are spin waves with two possible polarizations and an energy $\epsilon$ which
vanishes at small wavevectors $k$, $\epsilon = v k$. These spin waves correspond to local oscillations of $\boldsymbol{\Phi}$
about an orientation chosen by spontaneous breaking of the spin rotation symmetry in the N\'eel state, but which maintain low energy by fixing
the magnitude $|\boldsymbol{\Phi}|$. The spinwaves also interact weakly with each other, and the form of these
interactions can also be described by $\mathcal{S}_{\boldsymbol{\Phi}}$. All effects of these interactions are completely captured
by a single energy scale, $\rho_s$, which is the `spin stiffness', measuring the energy required to slowly twist the orientation of the N\'eel
order across a large spatial region. At finite temperatures, the thermal fluctuations of the interacting spin-waves can have strong
consequences. We will not describe these here (because they are purely consequences of classical thermal fluctuations),
apart from noting \cite{csy} 
that all these thermal effects can be expressed universally as functions of the dimensionless ratio $k_B T/\rho_s$.

For our future analysis, it is useful to have an alternative description of the low energy states above the N\'eel ordered state.
For the N\'eel state, this alternative description is, in a sense, a purely mathematical exercise: it does not alter any of the low
energy physical properties, and yields an {\em identical\/} low temperature theory for all observables when expressed
in terms of $k_B T /\rho_s$.
The key step is to express the vector field $\boldsymbol{\Phi}$ in terms of a $S=1/2$ complex spinor field $z_\alpha$,
where $\alpha = \uparrow \downarrow$ by
\begin{equation}
\boldsymbol{\Phi} = z_\alpha^\ast \boldsymbol{\sigma}_{\alpha \beta} z_\beta
\label{Phiz}
\end{equation}
where $\boldsymbol{\sigma}$ are the $2\times2$ Pauli matrices. Note that this mapping from $\boldsymbol{\Phi}$ to $z_\alpha$
is redundant. We can make a spacetime-dependent change in the phase of the $z_\alpha$ by the field $\theta(x,\tau)$
\begin{equation}
z_\alpha \rightarrow e^{i \theta} z_\alpha
\label{gauge}
\end{equation}
and leave $\boldsymbol{\Phi}$ unchanged. All physical properties must therefore also be invariant under Eq.~(\ref{gauge}),
and so the quantum field theory for $z_\alpha$ has a U(1) gauge invariance, much like that found in quantum electrodynamics.
The effective action for the $z_\alpha$ therefore requires introduction of an `emergent' 
U(1) gauge field $A_\mu$ (where $\mu = x, \tau$ is a 
three-component spacetime index). The field $A_\mu$ is unrelated the electromagnetic field, but is an internal
field which conveniently describes the couplings between the spin excitations of the antiferromagnet. As we have noted above, in the
N\'eel state, expressing the spin-wave fluctuations in terms of $z_\alpha$ and $A_\mu$ is a matter of choice, and the above theory
for the vector field $\boldsymbol{\Phi}$ can serve us equally well. The distinction between the two approaches
appears when we move out of the N\'eel state across quantum critical points into other phases (as we will see later): in some of these phases the emergent
$A_\mu$ gauge
field is no longer optional, but an essential characterization of the `quantum order' of the phase. As we did for $\mathcal{S}_{\boldsymbol{\Phi}}$,
we can write down the quantum field theory for $z_\alpha$ and $A_\mu$ by the constraints of symmetry and gauge invariance,
which now yields
\begin{equation}
\mathcal{S}_z =  \int d^2 r d \tau \biggl[
|(\partial_\mu -
i A_{\mu}) z_\alpha |^2 + s |z_\alpha |^2  + u (|z_\alpha |^2)^2 + \frac{1}{2e_0^2}
(\epsilon_{\mu\nu\lambda}
\partial_\nu A_\lambda )^2 \biggl] \label{Sz}
\end{equation}
For brevity, we have now used a ``relativistically'' invariant notation, and scaled away the spin-wave velocity $v$; the values
of the couplings $s,u$ are different from, but related to, those in $\mathcal{S}_{\boldsymbol{\Phi}}$. The Maxwell action for $A_\mu$ is generated from 
short distance $z_\alpha$ fluctuations, and it makes $A_\mu$ a dynamical field; its coupling $e_0$ is unrelated
to the electron charge. 
The action $\mathcal{S}_z$ is a valid description of the N\'eel state for $s<0$ (the critical upper value of $s$ will have fluctuation
corrections away from 0), where the gauge theory enters a Higgs phase with $\langle z_\alpha \rangle \neq 0$. This description of the N\'eel state as a Higgs phase has an analogy with the Weinberg-Salam theory of the weak interactions---in the latter case it is hypothesized that the condensation of a Higgs boson gives a mass to the $W$ and $Z$ gauge bosons, whereas here the condensation of $z_\alpha$ quenches the $A_\mu$ gauge boson.

\subsubsection{Triangular lattice}
\label{sec:triangle}

There have been numerous recent studies \cite{coldea} of the 
spin excitations of the insulator Cs$_2$CuCl$_4$. Just as in La$_2$CuO$_4$,
the dominant spin excitations are $S=1/2$ spins on the Cu ions, but now they reside on the vertices of a triangular lattice,
as sketched in Fig~\ref{neel}b. 
Such an antiferromagnet is well described by the Hamiltonian $H_0$, with a nearest neighbor
exchange $J$ and $i$ on the sites of the triangular lattice. From numerical studies of such spin systems \cite{bernu}, and also from observations \cite{coldea}
in Cs$_2$CuCl$_4$, the ground state of $H_0$ also has broken spin-rotation symmetry, but the pattern of spin polarization
is now quite different. We now replace Eq.~(\ref{sPhi}) by
\begin{equation}
\langle {\bf S}_j \rangle =  {\bf N}_1 \cos (K \cdot r_j) + {\bf N}_2 \sin (K \cdot r_j) \label{SN}
\end{equation}
where $r_i$ is the position of site $i$, and $K = (4 \pi /3a) (1, \sqrt{3})$ for the ordering pattern in Fig.~\ref{neel}b on a triangular
lattice of spacing $a$. The most important difference from Eq.~(\ref{sPhi}) is that we now require {\em two\/} 
orthogonal vectors ${\bf N}_{1,2}$ (${\bf N}_1 \cdot {\bf N}_2 =0$)
to specify the degenerate manifold of ground states. As for the square lattice, we can write down an effective action for ${\bf N}_{1,2}$
constrained only by the symmetries of the Hamiltonian. Minimization of such an action shows that the ordered state has
${\bf N}_1^2 = {\bf N}_2^2$ fixed to a value determined by parameters in the Hamiltonian, but are otherwise arbitrary.
Moving on to the analog of the spinor representation in Eq.~(\ref{Phiz}), we now
introduce another spinor $w_\alpha$, which parameterizes ${\bf N}_{1,2}$ by \cite{css}
\begin{equation}
{\bf N}_1 + i {\bf N}_2 = \varepsilon_{\alpha\gamma} w_\gamma \boldsymbol{\sigma}_{\alpha\beta} w_\beta,
\label{Nw}
\end{equation}
where $\varepsilon_{\alpha\beta}$ is the antisymmetric tensor.
It can be checked that $w_\alpha$ transforms as a $S=1/2$ spinor under spin rotations, and that under translations by a lattice vector $y$
$w_\alpha \rightarrow e^{-iK \cdot y/2} w_\alpha$. Apart from these global symmetries, we also have the analog of the gauge invariance
in Eq.~(\ref{gauge}). From the relationship of $w_\alpha$ to the physical observables in Eq.~(\ref{Nw}), we now find a $Z_2$ gauge
transformation 
\begin{equation}
w_\alpha \rightarrow \eta w_\alpha \label{z2w}
\end{equation}
where $\eta (r,\tau) = \pm 1$. This $Z_2$ gauge invariance will play an important role in the discussion in Section~\ref{sec:z2}. 
The low energy theory of the antiferromagnetically ordered state described by Eq.~(\ref{SN}) can now
be obtained from the effective action for the ${\bf N}_{1,2}$ or the $w_\alpha$. We won't write it out explicitly here, deferring it also to 
Section~\ref{sec:z2}

\subsection{Coupled dimer antiferromagnet}
\label{sec:dimer}

\begin{figure}[htbp]
  \centering
  \includegraphics[width=3in]{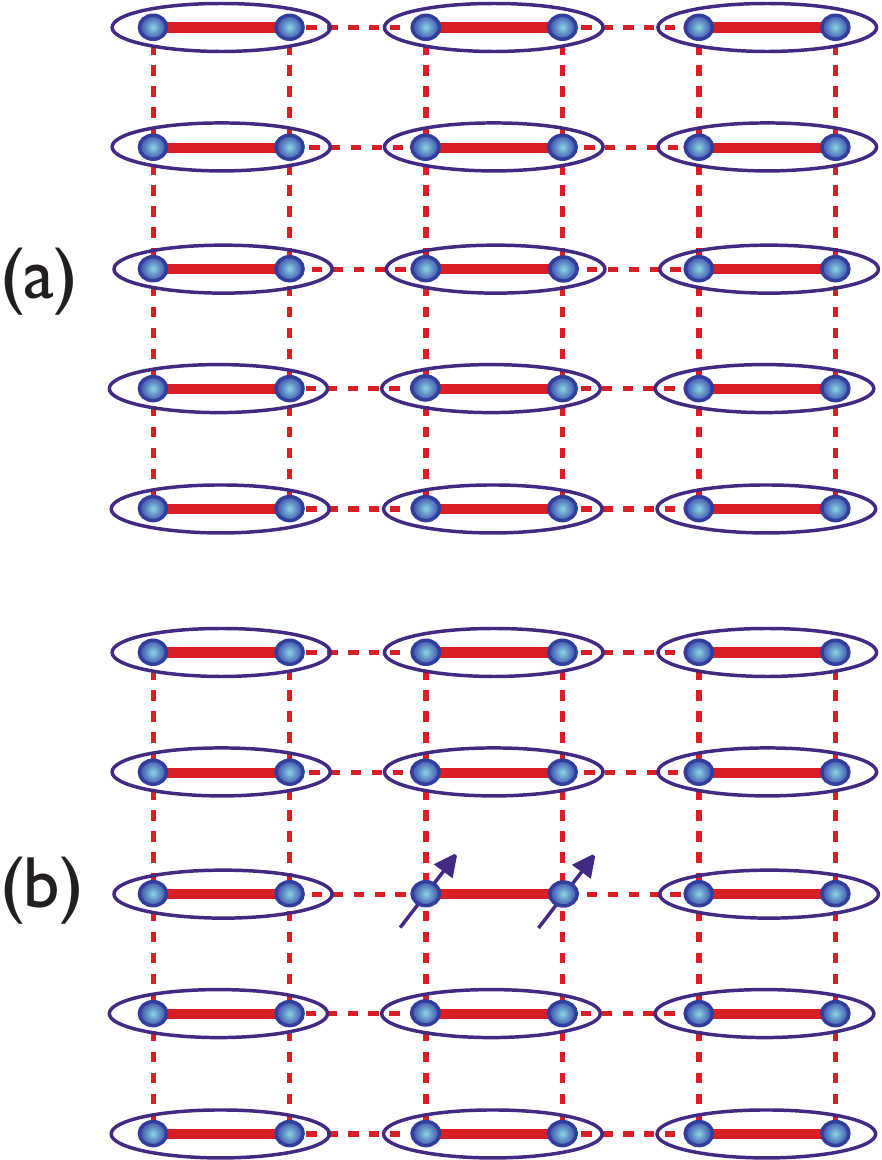}
  \caption{The coupled dimer antiferromagnet, described by the Hamiltonian $H_0$, with $J_{ij} = J$ on the full red
  lines, and $J_{ij} = J/g$ on the dashed red lines. (a) The large $g$ ground state, with each ellipse representing a singlet valence
  bond $(\left|\uparrow \downarrow \right\rangle - \left|\downarrow \uparrow \right\rangle)/\sqrt{2}$. 
  (b) The $S=1$ spin triplon excitation. The pair of parallel spins on the broken valence
  bond hops between dimers using the $J/g$ couplings.}
  \label{dimer}
\end{figure}

This spin model is illustrated in Fig.~\ref{dimer}. 
We begin with the square lattice antiferromagnet in Fig~\ref{neel}a,
and weaken the bonds indicated by the dashed lines to the value $J/g$. For $g=1$, this model reduces to the square lattice
model examined in Section~\ref{sec:neel}. For $g>1$, the model can be understood as a set of spin dimers, with the intra-dimer
exchange interaction $J$, and a weaker coupling between the dimers of $J/g$. A number of Cu compounds, such as TlCuCl$_3$
\cite{oosawa,ruegg}
and BaCuSi$_2$O$_6$ \cite{jaime}, realize such coupled
spin dimer models, although not in precisely the spatial pattern shown in Fig.~\ref{dimer}.

A simple argument shows that the large $g$ ground state is qualitatively distinct from the $g=1$ N\'eel state.
For $g=\infty$, the Hamiltonian separates into decoupled dimers, which have the obvious exact ground state
\begin{equation}
 \prod_{\langle ij \rangle
\in A} \frac{1}{\sqrt{2}} \left( \left| \uparrow \right\rangle_i
\left| \downarrow \right\rangle_j - \left| \downarrow
\right\rangle_i \left| \uparrow \right\rangle_j \right) \label{eq:dimer}
\end{equation}
where $A$ represents the set of links with full lines in Fig.~\ref{dimer}. This is the product of $S=0$ spin singlets on the dimers,
and so preserves full rotational invariance, unlike the N\'eel state. There is a gap towards $S=1$ excitations which are created by 
breaking the spin singlet valence bonds, as shown in Fig.~\ref{dimer}b. 
This gap ensures that perturbation theory in $1/g$ is well-defined,
and so the same structure is preserved for a finite range of $1/g$. The most significant change at non-zero $1/g$ is that the triplet
excitations becomes mobile, realizing a $S=1$ quasiparticle, sometimes called the triplon. Triplons have been clearly observed
in a number of neutron scattering experiments on spin dimer \cite{cavadini} and related \cite{collin} states.

It is interesting to understand this state and its triplon excitations from the perspective of quantum field theory. By the same procedure
as in Section~\ref{sec:neel}, we can argue that the theory in Eq.~(\ref{SP}) applies also in the present phase. However, we now need $s>0$ to 
preserve spin rotation invariance. The excitations of $\mathcal{S}_{\boldsymbol{\Phi}}$ about the $\boldsymbol{\Phi}=0$ saddle point
consists of the 3 polarizations of $\boldsymbol{\Phi}$ oscillations: we can identify these oscillations with a $S=1$ quasiparticle,
which clearly has all the characteristics of the triplon.

So we now have the important observation that $\mathcal{S}_{\boldsymbol{\Phi}}$ describes the $g=1$ N\'eel state for $s< s_c$ (the critical
value $s_c=0$ in mean field theory), and the large $g$ coupled dimer antiferromagnet for $s> s_c$. This suggest that there is
a quantum phase transition between these two states of $H_0$ at a critical value $g=g_c$ which is also described by $\mathcal{S}_{\boldsymbol{\Phi}}$. This is indeed the case, as we will discuss in Section~\ref{sec:lgw}.

\subsection{Valence bond solids}
\label{sec:vbs}

\begin{figure}[htbp]
  \centering
  \includegraphics[width=3in]{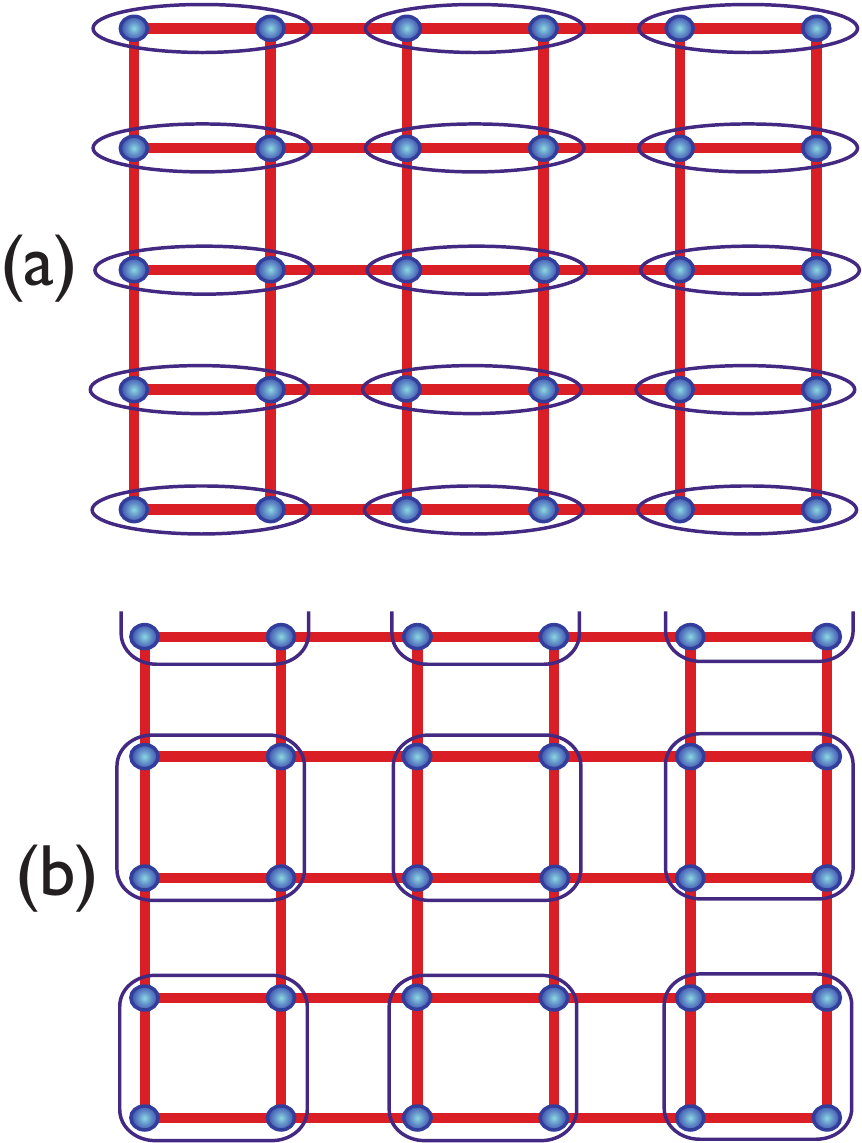}
  \caption{(a) Columnar VBS state of $H_0+H_1$ with $J_{ij} = J$ on all bonds. This state is the same as in Fig.~\ref{dimer}a, but the
  square lattice symmetry has been broken spontaneously. Rotations by multiples of $\pi/2$ about a lattice site yield the 4 degenerate states.
  (b) 4-fold degenerate plaquette VBS state with the rounded squares representing
  $S=0$ combination of 4 spins.}
  \label{vbs}
\end{figure}

The state in Eq.~(\ref{eq:dimer}), and its triplon excitations, are also acceptable caricatures of a  
valence bond solid (VBS) phase. However, the crucial difference is that the VBS appears as the ground state of
a Hamiltonian with full square lattice symmetry, and the space group symmetry is spontaneously broken, as shown in Fig.~\ref{vbs}a. 
Thus $H_0$ has all nearest-neighbor exchanges equal,
and additional exchange couplings, with full square lattice symmetries, are required to destroy the N\'eel order and restore
spin rotation invariance. Such VBS states, with spin rotation symmetry, but lattice symmetry spontaneously broken,
have been observed recently in the organic compound \cite{kato1} EtMe$_3$P[Pd(dmit)$_2$]$_2$
and in \cite{shlee} Zn$_x$Cu$_{4-x}$(OD)$_6$Cl$_2$. In the underdoped cuprates, bond-centered modulations in the local density of states observed in scanning tunneling microscopy experiments \cite{kohsaka} have a 
direct interpretation in terms of predicted doped VBS states \cite{poilblanc,vojta,rst2}.

We frame our discussion in the context of a specific perturbation, $H_1$, studied recently by Sandvik \cite{anders} and others \cite{rkk,shailesh}:
\begin{equation}
H_1 = -Q \sum_{\langle ijkl \rangle} \left( {\bf S}_i \cdot {\bf S}_j - \tfrac{1}{4} \right) \left( {\bf S}_k \cdot {\bf S}_l - \tfrac{1}{4} \right),
\end{equation}
where $\langle ijkl \rangle$ refers to sites on the plaquettes of the square lattice.
The ground state of the Hamiltonian $H_0 + H_1$ will depend upon the ratio $g=Q/J$. For $g=0$, we have the N\'eel ordered state of Section~\ref{sec:neel}. Recent numerical studies \cite{anders,rkk,shailesh} have shown convincingly that VBS order
is present for large $g$ (VBS order had also been found earlier in a related `easy-plane' model \cite{roger}). 
We characterize the VBS state by introducing a complex field, $\Psi$, whose expectation
value measures the breaking of space group symmetry:
\begin{equation}
\Psi = (-1)^{j_x} {\bf S}_j \cdot {\bf S}_{j+\hat{x}} + i  (-1)^{j_y} {\bf S}_j \cdot {\bf S}_{j+\hat{y}}. \label{defPsi}
\end{equation}
This definition satisfies the requirements of spin rotation invariance, and measures the lattice symmetry broken by the columnar VBS state because under a rotation about an even sublattice site by an angle $\phi=0,\pi/2,\pi,3\pi/2$ we have
\begin{equation}
\Psi \rightarrow e^{i \phi} \Psi \label{tPsi}
\end{equation}
Thus $\Psi$ is a convenient measure of the $Z_4$ rotation symmetry of the square lattice, and $\langle \Psi \rangle = 0$ in any state (such as the N\'eel state) in which this $Z_4$ symmetry is preserved.
From these definitions, we see that $\langle \Psi \rangle \neq 0$ in the VBS states; 
states with $\mbox{arg}( \langle \Psi \rangle ) = 0,\pi/2,\pi,3\pi/2$
correspond to the four degenerate VBS states with `columnar' dimers in Fig.~\ref{vbs}a, while states with 
$\mbox{arg}( \langle \Psi \rangle ) = \pi/4,3\pi/4,5\pi/4,7\pi/4$ correspond to the plaquette VBS states in Fig.~\ref{vbs}b. 

\begin{figure}[htbp]
  \centering
  \includegraphics[width=2.5in]{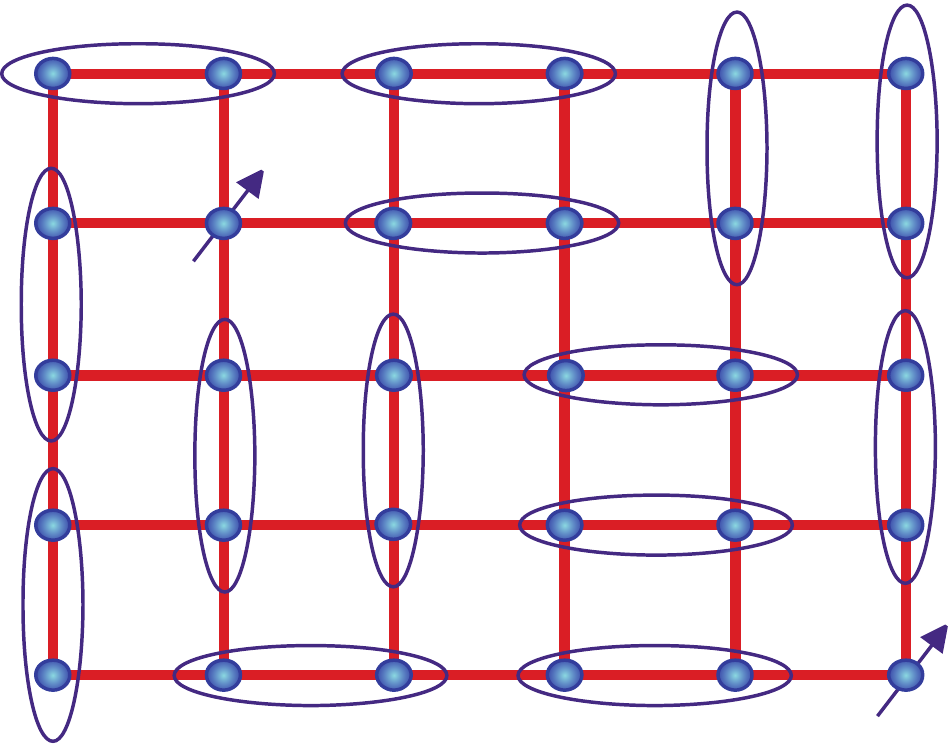}
  \caption{Caricature of a spin liquid state. The valence bonds are entangled between different pairings of the spins, only one
  of which is shown. Also shown are two unpaired $S=1/2$ spinons, which can move independently in the spin liquid
  background.}
  \label{rvb}
\end{figure}

VBS order with $\langle \Psi \rangle \neq 0$ was predicted in early theoretical work \cite{rsl} on the destruction
of N\'eel order on the square lattice. We now present the arguments, 
relying on the theory $\mathcal{S}_z$ in Eq.~(\ref{Sz}) as a representation
of the excitations of the N\'eel state. We mentioned earlier that $\mathcal{S}_z$ describes the N\'eel state for $s<s_c$
(with $s_c=0$ in mean field theory). 
Let us assume that $\mathcal{S}_z$ remains a valid description across a quantum phase transition at $s=s_c$,
and ask about the nature of the quantum state \cite{mv} for $s>s_c$. A conventional fluctuation analysis about the $z_\alpha=0$ saddle point reveals two types of excitations: ({\em i\/}) the $S=1/2$ $z_\alpha$ quanta which realize
bosonic quasiparticles known as spinons (see Fig~\ref{rvb}), and ({\em ii\/}) the $A_\mu$ `photon' which represents
a spinless excitations we can identify with the resonance between different valence bond configurations \cite{fk}. 
More precisely, the $A_\mu$ `magnetic flux' determines the relative phases of
different possible pairings of the electrons
into valence bonds. It would also appear that the $s>s_c$ phase of $\mathcal{S}_z$ breaks no symmetries of the underlying spin Hamiltonian $H_0+ H_1$. However, there are well-known arguments \cite{polyakov}
that this phase of $\mathcal{S}_z$ breaks a `hidden' or `topological' symmetry. Let us define the topological current by
\begin{equation}
J_\mu = \epsilon_{\mu\nu\lambda} \partial_\nu A_\lambda \label{JA}
\end{equation}
where $\epsilon_{\mu\nu\lambda}$ is the totally antisymmetric tensor in 3 spacetime dimensions,
so that $J_t$ is the `magnetic' flux. The conservation law, $\partial_\mu J_\mu = 0$, is a trivial consequence of this definition, and represents the conservation of $A_\mu$ flux
in the $s>s_c$ phase of $\mathcal{S}_c$. However, by defining 
\begin{equation}
J_\mu = \partial_\mu \zeta, \label{Jchi}
\end{equation}
where $\zeta$ is a free massless scalar field at low energies, we can also identify $J_\mu$ as the Noether current associated with the shift symmetry $\zeta \rightarrow \zeta + \mbox{constant}$. This shift symmetry is spontaneously broken in the $s>s_c$ phase
of $\mathcal{S}_z$, and $\zeta$ is the associated Goldstone boson. (Indeed, this Goldstone boson is, of course, just the
$A_\mu$ photon which has only a single allowed polarization in 2+1 dimensions.) Nevertheless, these observations appear purely formal because they only involve a restatement of the obvious divergencelessness of the flux in Eq.~(\ref{JA}), and the shift symmetry appears unobservable.

The key point made in Ref.~\onlinecite{rsl} was that the above shift symmetry is, in fact, physically observable,
and is an enlargment of the $Z_4$ rotation symmetry of the square lattice. They focused attention on the 
monopole operator, $V$, which changes the total $A_\mu$ flux by $2 \pi$ (by the Dirac argument, because total flux is conserved,
any such tunneling event must be accompanied by a thin flux tube carrying flux $2\pi$---this string is unobservable to the $z_\alpha$
quanta which carry unit $A_\mu$ charge). From Eq.~(\ref{Jchi}), the flux density is measured by $\partial_t \zeta$, and so
its canonically conjugate variable under the Maxwell action in Eq.~(\ref{Sz}) is $\zeta/e_0^2$; consequently, the operator which shifts flux by $2 \pi$ is  
\begin{equation}
V = \exp \left(i \frac{2 \pi \zeta}{e_0^2} \right).
\end{equation}
In terms of the underlying spin model, this monopole is a tunneling
event between semiclassically stable spin textures which differ in their `skyrmion' number. 
By a careful examination of the phase factors \cite{haldane} 
associated
with such tunnelling events, it is possible to show that $V$ transforms just like $\Psi$ in Eq.~(\ref{tPsi})
under a lattice $Z_4$ rotation (and also under other square lattice space group operations). 
However, this is also the action of $V$ under the shift symmetry, and so we may
identify the $Z_4$ rotation with a particular shift operation. 
We may also expect that there are additional
allowed terms that we may add to $\mathcal{S}_z$ which reduce the continuous shift symmetry to 
only a discrete $Z_4$ symmetry, and this is indeed the case (see Eq.~(\ref{V4})). With this reduction
of a continuous to a discrete symmetry, we expect that the Goldstone boson (the photon) will acquire a small gap,
as will become clearer in Section~\ref{sec:dcp}. 
These arguments imply that we may as well {\em identify\/} the two operators \cite{rsl,senthil2}:
\begin{equation}
V \sim \Psi \label{VPsi}
\end{equation}
with a proportionality constant that depends upon microscopic details. The breaking of shift symmetry
in the $s>s_c$ phase of $\mathcal{S}_z$ now implies that we also have $\langle \Psi \rangle \neq 0$,
and so reach the remarkable conclusion that VBS order is present in this phase. 
Thus it appears that $\mathcal{S}_z$ describes
a transition from a N\'eel to a VBS state with increasing $s$, and this transition will be discussed further in
Section~\ref{sec:dcp}. We will also present there a comparison of the numerical observation of VBS order in $H_0+H_1$
with present theory.

\subsection{$Z_2$ spin liquids}
\label{sec:z2} 

The discussion in Section~\ref{sec:vbs} on the entanglement between valence bonds illustrated
in Fig.~\ref{rvb} raises a tantalizing possibility. Is it possible that this resonance is such that we obtain
a ground state of an antiferromagnet which preserves {\em all\/} symmetries of the Hamiltonian ?
In other words, we restore the spin rotation invariance of the N\'eel state but do not break any symmetry 
of the lattice, obtaining a state called resonating valence bond (RVB) liquid. Such a state would be a
Mott insulator in the strict sense, because it has an odd number of electrons per unit cell, and so cannot
be smoothly connected to any band insulator.
For phases like those described by $\mathcal{S}_z$, with gapped spinons and a U(1) photon,
the answer is no: the monopole operator $V$, identified with the VBS order $\Psi$, has long-range correlations
in a state with a well-defined U(1) photon.

The remainder of the next two sections will discuss a variety of approaches in which this monopole instability can
be suppressed to obtain RVB states.

The route towards obtaining a RVB state in two dimensions
was presented in Refs.~\onlinecite{rst1,sstri,wen}. The idea is to use the Higgs mechanism to 
reduce the unbroken gauge invariance from U(1) to a discrete gauge group, $Z_2$, and so reduce the strength
of the gauge-flux fluctuations. To break U(1) down to $Z_2$, we need a Higgs scalar, $\Lambda$, 
that carries U(1) charge 2, so that $\Lambda \rightarrow e^{2i\theta} \Lambda$ under the transformation
in Eq.~(\ref{gauge}). Then a phase with $\langle \Lambda \rangle \neq 0$ would break the U(1) symmetry, in
the same manner that the superconducting order parameter breaks electromagnetic gauge invariance. 
However, a gauge transformation with $\theta = \pi$, while acting non-trivially on the $z_\alpha$, would leave
$\Lambda$ invariant: thus there is a residual $Z_2$ gauge invariance which constrains the structure
of the theory in the Higgs phase. 

What is the physical interpretation of the field $\Lambda$, and how does its presence characterize the resulting
quantum state {\em i.e.\/} what are the features of this $Z_2$ RVB liquid which distinguishes it from other quantum
states ? This is most easily determined by writing down the effective action for $\Lambda$, constrained only
by symmetry and gauge invariance, including its couplings to $z_\alpha$. In this manner, we expand the theory
$\mathcal{S}_z$ to $\mathcal{S}_z + \mathcal{S}_\Lambda$, with
\begin{equation}
\mathcal{S}_\Lambda =  \int d^2 r d \tau \left[
|(\partial_\mu - 2 i A_\mu) \Lambda_a |^2 + \tilde{s} |\Lambda_a |^2 + \tilde{u} |\Lambda_a|^4 -i 
\Lambda_a \varepsilon_{\alpha\beta} z_\alpha^\ast \partial_a z_\beta^\ast + \mbox{c.c.} \right] \label{SL}
\end{equation}
We have introduced multiple fields $\Lambda_a$, with a spatial index $a$, which is necessary to account
for the space group symmetry of the underlying lattice---this is a peripheral complication which we will gloss over here.
The crucial term is the last one coupling $\Lambda_a$ and $z_\alpha$: it indicates that $\Lambda$ is a molecular state of a pair of spinons in a spin-singlet state; this pair state has a ``$p$-wave'' structure, as indicated by the spatial gradient $\partial_a$. 
It is now useful to examine the mean field phase diagram of $\mathcal{S}_z
+ \mathcal{S}_\Lambda$, as a function of the two ``masses'' $s$ and $\tilde{s}$. We have 2 possible condensates,
and hence 4 possible phases \cite{rst2}:\\
({\em i\/}) $s<0$, $\tilde{s}>0$: This state has $\langle z_\alpha \rangle \neq 0$ and $\langle \Lambda \rangle =0$.
We may ignore the gapped $\Lambda$ modes, and this is just the N\'eel state of Section~\ref{sec:neel}.\\
({\em ii\/}) $s>0$, $\tilde{s}>0$: This state has $\langle z_\alpha \rangle = 0$ and $\langle \Lambda \rangle  = 0$.
Again, we may ignore the gapped $\Lambda$ modes, and this is the VBS state of Section~\ref{sec:vbs}.\\
({\em iii\/}) $s<0$, $\tilde{s}<0$: This state has $\langle z_\alpha \rangle \neq 0$ and $\langle \Lambda \rangle  \neq 0$. Because of the $z_\alpha$ condensate, this state breaks spin rotation invariance, and we determine the spin configuration by finding the lowest energy $z_\alpha$ mode in the background of a non-zero $\langle \Lambda \rangle$ in 
Eq.~(\ref{SL}), which is
\begin{equation}
z_\alpha = \left( w_\alpha e^{i \langle \Lambda \rangle \cdot r} + \varepsilon_{\alpha\beta} w_\beta^\ast e^{-i \langle
\Lambda \rangle \cdot r} \right)/\sqrt{2}, \label{zw}
\end{equation}
with $w_\alpha$ a constant spinor.
Inserting Eq.~(\ref{zw}) into Eq.~(\ref{Phiz}) we find that $\boldsymbol{\Phi}$ is space-dependent
so that $\langle {\bf S}_i \rangle$ obeys Eq.~(\ref{SN}) with ${\bf N}_{1,2}$ given by Eq.~(\ref{Nw}) and 
the wavevector $K = (\pi,\pi) + 2\langle \Lambda \rangle$.
Thus this state is the same as the coplanar spin-ordered state of Section~\ref{sec:triangle} !
The $Z_2$ gauge transformation in Eq.~(\ref{z2w}) is the same as the $Z_2$ $\theta=\pi$
transformation we noted earlier in this subsection.\\
({\em iv\/}) $s>0$, $\tilde{s}<0$: This state has $\langle z_\alpha \rangle = 0$ and $\langle \Lambda \rangle  \neq 0$.
This is the $Z_2$ spin liquid (or Higgs) state we are after. Spin rotation invariance is preserved, and there is no VBS order 
because monopoles are suppressed
by the $\Lambda$ condensate, $\langle \Psi \rangle \sim \langle V \rangle = 0$.

We are now in a position to completely characterize this $Z_2$ spin liquid. One class of excitations of this state
are the $z_\alpha$ spinons which now carry a unit $Z_2$ ``electric'' charge. A second class are vortices in which the
phase $\Lambda$ winds by $ 2 \pi$ around a spatial point: these are analogous to Abrikosov vortices 
in a BCS superconductor, and so carry $A_\mu$ flux $\pi$. Further, because the $A_\mu$ flux can be 
changed by $2 \pi$ by monopoles, only the flux modulo $2 \pi$ is significant. Consequently there is only
a single type of vortex, which is often called a vison. Two visons can annihilate each other, and so we can
also assign a $Z_2$ ``magnetic'' charge to a vison. Finally, from the analog of the Abrikosov solution
of a vortex, we can deduce that a $Z_2$ electric charge picks up a factor of $(-1)$  when transported
around a vison. This $Z_2 \times Z_2$ structure is the fundamental characterization of 
the quantum numbers characterizing this spin liquid, and its electric
and magnetic excitations. An effective lattice gauge theory (an `odd' $Z_2$ gauge theory) 
of these excitations, and of the transition
between this phase and the VBS state was described in Refs.~\onlinecite{jalabert} (a detailed derivation
was published later \cite{vs}).
We also note that $Z_2 \times Z_2$ is the simplest of a large class of topological
structures described by Bais and collaborators \cite{bais1,bais2} in their work on Higgs phases of discrete gauge theories.

Since the work of Refs.~\onlinecite{rst1,sstri,wen}, the same $Z_2 \times Z_2$ structure has
been discovered in a variety of other models. The most well-known is the toric code of Kitaev \cite{kitaev},
found in an exactly solvable $Z_2$ gauge theory, which has found applications as a topologically
protected quantum memory \cite{preskill}. An exactly solvable model of $S=1/2$ spins in a $Z_2$ spin liquid
phase was also constructed by Wen \cite{wen2}. Senthil and Fisher \cite{sf} have described $Z_2$ spin liquid
states in easy-plane magnets and superconductors. Sondhi and Moessner \cite{sondhi} have studied
quantum dimer models which also exhibit $Z_2$ spin liquid phases. Finally, Freedman {\em et al.} \cite{freedman}
have presented a classification of topological quantum phases using Chern-Simons gauge theory
in which the $Z_2 \times Z_2$ state is the simplest case.

\section{Quantum critical points and phases}
\label{sec:qcp}

In Section~\ref{sec:ins}, we saw several examples of insulating quantum magnets
which were tuned between two distinct quantum phases by a coupling $g$. 
We will now turn our attention to the transition between these phases, and show that the critical
points realize non-trivial quantum states, as was promised in Section~\ref{sec:intro}.
In Section~\ref{sec:dcp} we will also describe interesting cases where such critical quantum states
are found not just at isolated points in the phase diagram, but extend into a finite critical quantum phase.

\subsection{Landau-Ginzburg-Wilson criticality}
\label{sec:lgw}

We begin by considering the coupled dimer antiferromagnet of Section~\ref{sec:dimer}. With increasing $g$,
this model has a phase transition from the N\'eel state in Fig.~\ref{neel}a to the dimer state in Fig,~\ref{dimer}a.
We also noted in Section~\ref{sec:dimer} that this transition is described by the theory $\mathcal{S}_{\boldsymbol{\Phi}}$
in Eq.~(\ref{SP}) by tuning the coupling $s$. From numerical studies \cite{matsumoto}, we know that the transition in the dimer
antiferromagnet occurs at $g=1.9107\dots$, and the measured critical exponents are in excellent agreement with those
expected from $\mathcal{S}_{\boldsymbol{\Phi}}$. The quantum state at $g=g_c$ is described by a conformal field 
theory (CFT), a scale-invariant structure of correlations which describes a renormalization group fixed point (known
as the Wilson-Fisher fixed point)
of $\mathcal{S}_{\boldsymbol{\Phi}}$. The ground state breaks no symmetries, and there is a spectrum
of gapless excitations whose structure is not completely understood: the spectrum is partly characterized by the scaling dimensions
of various operators in the CFT, which can be computed either numerically or using a number of perturbative 
field-theoretic methods.

We place this critical quantum state in the category of Landau-Ginzbug-Wilson (LGW) criticality, because
the above understanding followed entirely from an identification of the broken symmetry in the N\'eel phase
as characterized by the field $\boldsymbol{\Phi}$, and a subsequent derivation of the effective action for
$\boldsymbol{\Phi}$ using only symmetry to constrain the couplings between the long-wavelength modes. 
This procedure was postulated by LGW for classical thermal phase transitions, and it works successfully for the present
quantum transition.

Striking experimental evidence for this picture has been found recently \cite{ruegg2} in the coupled-dimer antiferromagnet
TlCuCl$_3$. One of the primary predictions of $\mathcal{S}_{\Phi}$ is that the low-lying excitations
evolve from 2 spin wave modes in the N\'eel state, to the 3 gapped triplon particles in the dimer state.
The extra mode emerges as a longitudinal spin excitation near the quantum critical point, and this
mode has been observed in neutron scattering experiments.

\subsection{Critical U(1) spin liquids}
\label{sec:dcp}

Consider the phase transition in the model $H_0+H_1$, as a function of the ratio $g=Q/J$,
in a model which preserves full square lattice symmetry. The crucial difference from the model
of Section~\ref{sec:lgw} is that there is only one $S=1/2$ spin per unit cell, and so there is no
trivial candidate for a $S=0$ ground state (like that in Eq.~(\ref{eq:dimer})) in the non-N\'eel phase. Indeed, this
argument is sufficient to exclude $\mathcal{S}_{\boldsymbol{\Phi}}$ from being the appropriate
quantum field theory for the transition, because $\mathcal{S}_{\boldsymbol{\Phi}}$ has a trivial
gapped ground state for $s$ large.

Instead, as we discussed in Sections~\ref{sec:neel} and~\ref{sec:vbs}, the ground state of $H_0+H_1$ evolved with increasing $Q/J$ from the N\'eel state in Fig.~\ref{neel}a to the VBS states in Fig.~\ref{vbs}. We also argued that
the low energy theory for the vicinity of this transition was given by the theory $\mathcal{S}_z$ in Eq.~(\ref{Sz}) of $S=1/2$
spinons coupled to a U(1) gauge field $A_\mu$. The N\'eel order parameter $\boldsymbol{\Phi}$ was related
to the spinons by Eq.~(\ref{Phiz}), while the monopole operator for the $A_\mu$ gauge field was identified
with the VBS order parameter $\Psi$ in Eq.~(\ref{VPsi}). Notice that this low energy effective theory
is not expressed directly in terms of either the N\'eel order $\boldsymbol{\Phi}$ or the VBS order $\Psi$. This proposal therefore
runs counter to the LGW procedure, which would lead directly for a theory of coupled $\boldsymbol{\Phi}$, $\Psi$ fields, like those
presented in Ref.~\onlinecite{fisherliu}. Instead, the low energy theory involves an emergent U(1) gauge field $A_\mu$.

Now we ask about the fate of this U(1) gauge field across the N\'eel-VBS transition. We noted in Section~\ref{sec:neel}
that $A_\mu$ was quenched by the Higgs phenomenon in the N\'eel state, while in Section~\ref{sec:vbs} we pointed
out that $A_\mu$ was gapped in the VBS state by the reduction of the continuous shift symmetry of the field $\zeta$
in Eq.~(\ref{Jchi}) to a $Z_4$ lattice rotation symmetry. However, is it possible that $A_\mu$ becomes gapless
and {\em critical\/} in the transition from the N\'eel to the VBS state ?

\begin{figure}[htbp]
  \centering
  \includegraphics[width=3.5in]{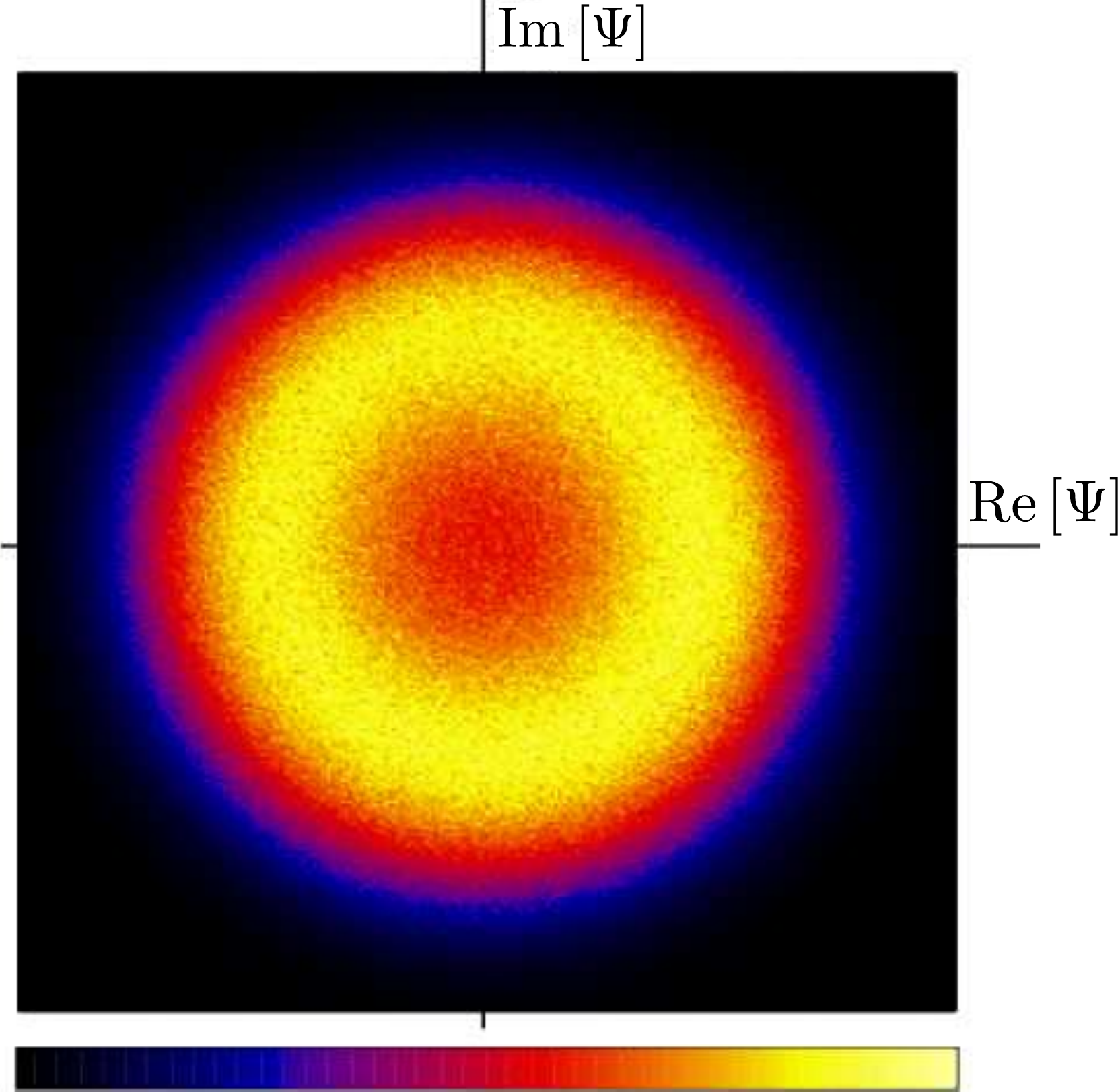}
  \caption{Histogram of the VBS order parameter defined in Eq.~(\ref{defPsi}) in the numerical
  study \cite{anders} by Sandvik of $H_0+H_1$. The circular symmetry is emergent and implies a Goldstone boson
  which is the emergent photon $A_\mu$.}
  \label{sandvik}
\end{figure}

An affirmative answer to the last question would yield a most interesting U(1) spin liquid state \cite{senthil1,senthil2}.
For $A_\mu$ to remain gapless, it is required that the continuous shift symmetry in $\zeta$ be preserved.
The connection with the $Z_4$ lattice rotation symmetry implies that the resulting U(1) spin liquid
state actually has a continuous spatial rotation symmetry. In other words, the wavefunction describing the 
resonating valence bonds has an emergent circular symmetry, even though the spins on the valence
bonds reside on the sites of a square lattice. 
Recent numerical studies by Sandvik \cite{anders}
and others \cite{shailesh} on the model $H_0+H_1$ have observed this remarkable phenomenon. 
They computed the histogram in their simulation for the VBS order $\Psi$ not too far from the transition,
and the results are shown in Fig.~\ref{sandvik}.
The most striking feature is the almost perfect circular symmetry of the histogram; similar results have
also been obtained in Ref.~\onlinecite{shailesh}, and earlier for an easy-plane model \cite{roger}.  This
circular symmetry also means that the system
has not made a choice between the VBS orders in Figs.~\ref{vbs}a and~\ref{vbs}b, for which the distribution in Fig.~\ref{sandvik} 
would be oriented along
specific compass directions (N,E,S,W, and NE,SE,SW,NW respectively). These numerical results strongly support
our claim that the theory $\mathcal{S}_z$ is the proper low energy theory for the transition from the N\'eel
to the VBS state, rather than the analog of the theory in Ref.~\onlinecite{fisherliu}.

Let us restate the emergence of circular symmetry in  
more formal terms. The theory $\mathcal{S}_z$ has a global SU(2) spin rotation symmetry, and also a global U(1) symmetry associated with the rotation in phase of the monopole $V$ due to the shift symmetry in massless
scalar $\zeta$ (noted below Eq.~(\ref{Jchi})). The underlying spin model $H_0+H_1$ also has a global SU(2) spin rotation
symmetry, but only a $Z_4$ lattice rotation symmetry. 
The simplest term we can add to $\mathcal{S}_z$ which breaks its U(1) symmetry down to $Z_4$ is
\begin{equation}
\mathcal{S}_V = \lambda \int d^2r d\tau \left[ V^4 + \mbox{c.c.} \right], \label{V4}
\end{equation}
and a non-zero $\lambda$ gives a mass to the scalar $\zeta$. The circular symmetry of Fig~\ref{sandvik} implies that
$\lambda$
scales under renormalization to small values \cite{csy,senthil2} over the length scales studied by the simulation.

A separate question is whether the N\'eel-VBS transition is second-order, with a critical point where $\lambda$
scales all the way to zero.
Numerical results are inconclusive at present \cite{anders,rkk,shailesh,kuklov,flavio}, with indications of both a
second-order and a weakly first-order transition. In any case, 
it is already clear from the present results (see Fig~\ref{sandvik})
that there is a substantial length scale, much larger than the spin correlation length, over which monopoles described
by Eq.~(\ref{V4}) are suppressed, and the spin liquid theory of $z_\alpha$ spinons and the $A_\mu$ gauge field applies. Indeed, the length scale at which $\zeta$ acquires a mass, and the
$z_\alpha$ spinons are confined, is larger than all system sizes studied so far.

A separate class of critical U(1) spin liquids appear in models in which the spinons are represented as fermions.
These states of antiferromagnets were proposed by Rantner and Wen \cite{rantwen}, building upon early work of
Affleck and Marston \cite{brad}, and a complete theory was presented by Hermele {\em et al.} \cite{stableu1,motherasl}. 
They are closely analgous to the U(1) spin liquid discussed above
but with two important differences: ({\em i\/}) the bosonic $z_\alpha$ spinons are replaced by massless Dirac fermions
$\psi_\alpha$, and ({\em ii\/}) they are postulated to exist not at isolated critical points between phases
but as generic critical quantum phases. The latter difference requires that the CFT of these spin liquids has no relevant
perturbations which are permitted by the symmetries of the underlying antiferromagnet. This has so far only been established
in a $1/N$ expansion in models with SU($N$) spin symmetry. These spin liquids have also been proposed \cite{aslkagome} as
candidate ground states for the kagome lattice antiferromagnet observed recently in Zn$_x$Cu$_{4-x}$(OH)$_6$Cl$_{2}$.

\section{Superfluids and Metals}
\label{sec:ms}

Our discussion of quantum phases and critical points has so far been limited to examples
drawn from quantum magnetism, in which the primary degrees of freedom are stationary
unpaired electrons which carry spin $S=1/2$. We have seen that a
plethora of interesting quantum states are possible, many of which have found realizations in experimental
systems. In this section we will widen the discussion to including systems which have also
allow for motion of mass and/or charge. Thus in addition to insulators, we will find states
which are metals or superfluids/superconductors. We will find that many phases and transitions
can be understood by mappings and analogies from the magnetic systems.

The simplest examples of a quantum phase transition involving motion of matter are
found in systems of trapped ultracold atoms, which have been reviewed earlier \cite{bloch}.
Here the degrees of freedom are neutral atoms, and so the motion involves mass, but not
charge, transport. Bosonic atoms are placed in a optical lattice potential created by standing waves of laser
light, and their ground state is observed as a function of the depth of the periodic potential. 
For weak potential, the ground state is a superfluid, while for a strong potential the ground state may
be an insulator.

A crucial parameter determining the nature of the insulator, and of the superfluid-insulator
transition, is the filling fraction, $f$, of the optical lattice. This is the ratio of number of bosons
to the number of unit cell of the optical lattice. 

\begin{figure}[htbp]
  \centering
  \includegraphics[width=2.in]{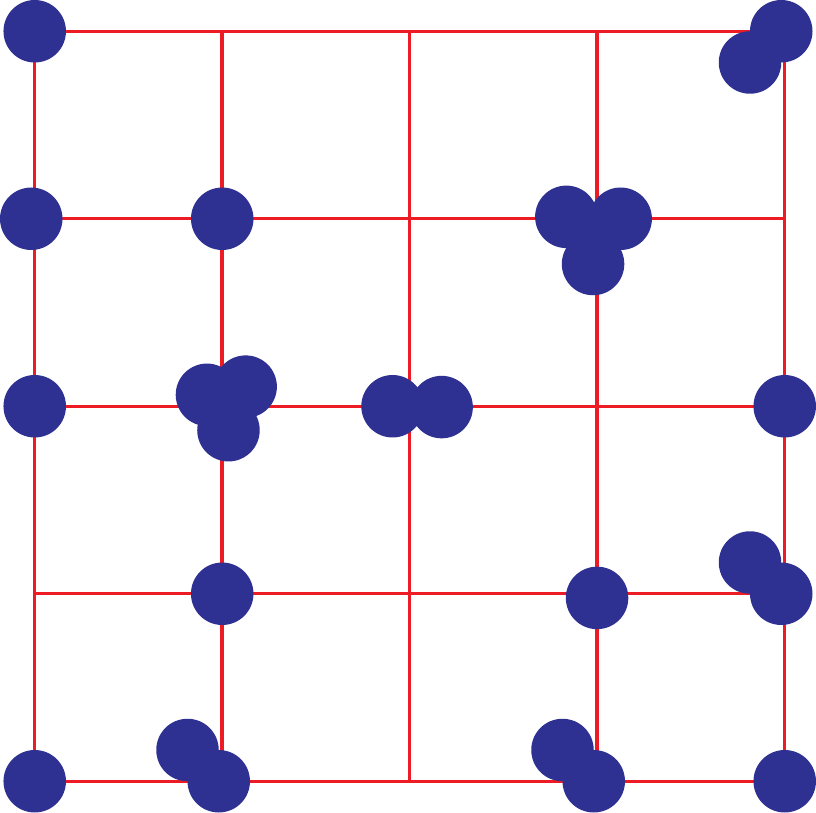}\\~\\
  \includegraphics[width=2in]{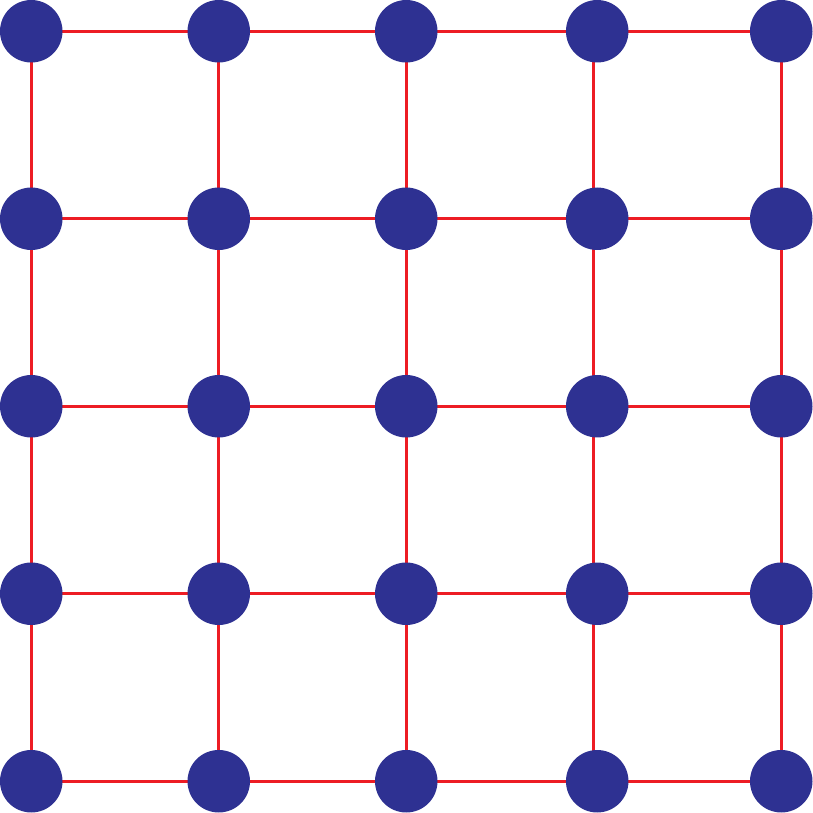}
  \caption{Superfluid (top) and insulating states of bosons with repulsive interactions on a square
  lattice at $f=1$.}
  \label{integer}
\end{figure}

We begin by considering the simplest case, $f=1$, which is illustrated in Fig.~\ref{integer}.
For a weak potential, the bosons move freely through the lattice, and form a Bose-Einstein condensate, leading to superfluidity.
In the site representation shown in Fig.~\ref{integer}, the superfluid state is a quantum superposition of many particle
configurations, only one of which is shown. The configurations involve large fluctuations in the number of particles
on each site, and it is these fluctuations that are responsible for easy motion of mass currents, and hence superfluidity.
For a strong potential, the bosons remain trapped in the lattice wells, one each per unit cell, forming the simple `Mott insulator'
shown in Fig.~\ref{integer}. We can describe quantum phase transition between these states by using
the LGW order-parameter approach, just as we did for the coupled-dimer antiferromagnet in Section~\ref{sec:lgw}. The order
parameter now is the condensate wavefunction, which is represented by a complex field $\Phi$. Indeed, this field has
2 real, components but it is otherwise similar to the 3-component field $\boldsymbol{\Phi}$ considered in Section~\ref{sec:neel}.
The LGW theory for the superfluid-insulator transition \cite{fwgf} is then given by action in Eq.~(\ref{SP}), with the replacement
$\boldsymbol{\Phi} \rightarrow \Phi$.

\begin{figure}[htbp]
  \centering
  \includegraphics[width=2in]{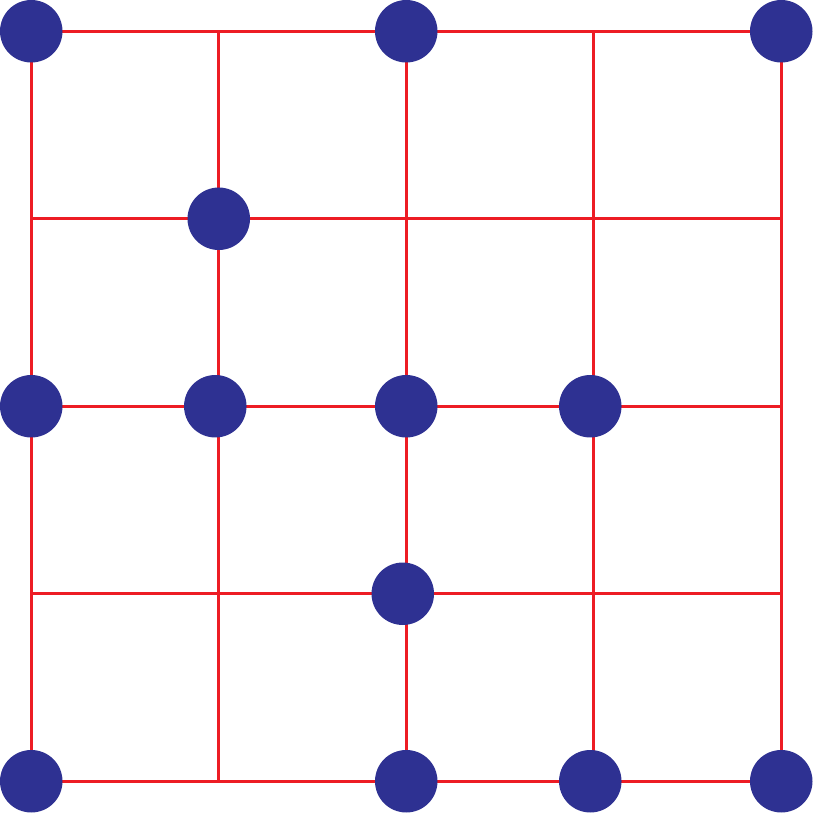}\\~\\
  \includegraphics[width=4.3in]{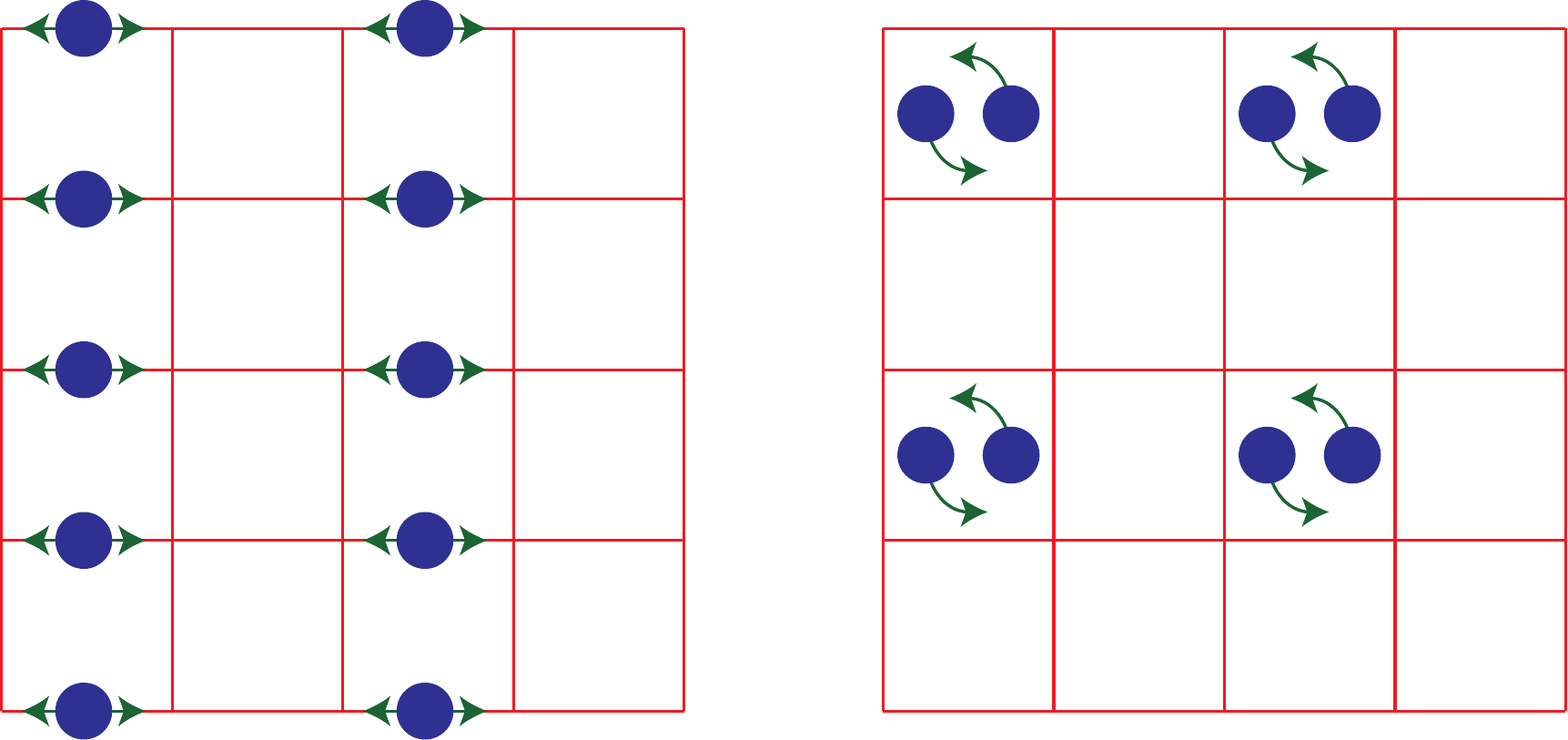}
  \caption{Superfluid (top) and two possible insulating states of bosons on a square
  lattice at $f=1/2$. The single bosons on the links in the first insulator resonate between the two sites on the ends
  of the link, and this insulator is similar to the VBS state in Fig.~\ref{vbs}a. Similarly the two bosons on the plaquette in the second insulator resonate around the plaquette, and this insulator is similar to the VBS state in Fig.~\ref{vbs}b.}
  \label{half}
\end{figure}

Now let us move to $f=1/2$, which is illustrated in Fig.~\ref{half}.
The superfluid state is fairly similar to that at $f=1$, with large number fluctuations on each site. 
However, the nature of the insulator is now very different. Indeed, in a single-band ``boson Hubbard model'' with
purely on-site repulsive interactions between the particles, the ground state remains a superfluid even at infinitely
large repulsion (`hard-core' bosons) : this is clear from Fig.~\ref{half}, where it is evident that number fluctuations are possible at site
even when double occupancy of the sites is prohibited. However, once off-site interactions are included, interesting insulating
states become possible. The nature of these phases can be understood by a very useful mapping between hard-core bosons and the $S=1/2$ 
antiferromagnets studied in Sections~\ref{sec:neel} and~\ref{sec:vbs}; the Hilbert spaces of the two models are seen to be identical
by identifying an empty site with a down spin (say), and site with a boson with an up spin. Further, in this mapping, an easy-plane N\'eel state
of Section~\ref{sec:neel} maps onto the superfluid state of the bosons. It is then natural to ask for the fate of the VBS states
in Fig.~\ref{vbs} under the mapping to the hard-core bosons. A simple analysis yields the insulating states
shown in Fig.~\ref{half}. The same reasoning also allows us to analyze the superfluid-insulator transition for $f=1/2$ bosons, by connecting
it to the N\'eel-VBS transition in Section~\ref{sec:dcp}. In this manner, we conclude that the low energy quantum states
near the superfluid-insulator transition are described by the U(1) gauge field theory $\mathcal{S}_z$ in Eq.~(\ref{Sz});
the only change is that an additional `easy-plane' term is permitted with lowers the global symmetry from SU(2) to U(1). 
The $z_\alpha$ quanta of this theory are now emergent `fractionalized' bosons which carry boson number 1/2 
(in units in which the bosons illustrated
in Fig.~\ref{half} carry boson number 1); a microscopic picture of these fractionalized bosons can be obtained by applying
the above mapping to the spinon states in Fig.~\ref{rvb}.

The crucial result we have established above is that LGW criticality does not apply at $f=1/2$. By extensions
of the above argument it can be shown that LGW criticality applies only at {\em integer\/} values of $f$. Superfluid-insulator
transitions at fractional $f$ are far more complicated.

The above gauge theory for bosons at $f=1/2$ has been extended to a wide variety of cases,
including a large number of other filling fractions \cite{bbbss}, and models of paired electrons on the square lattice \cite{lfs,balents3}
also at general filling fractions. The resulting field theories are more complicated that $\mathcal{S}_z$, but all involve
elementary degrees of freedom which carry fractional charges coupled to emergent U(1) gauge fields.

\begin{figure}[htbp]
  \centering
  \includegraphics[width=3.5in]{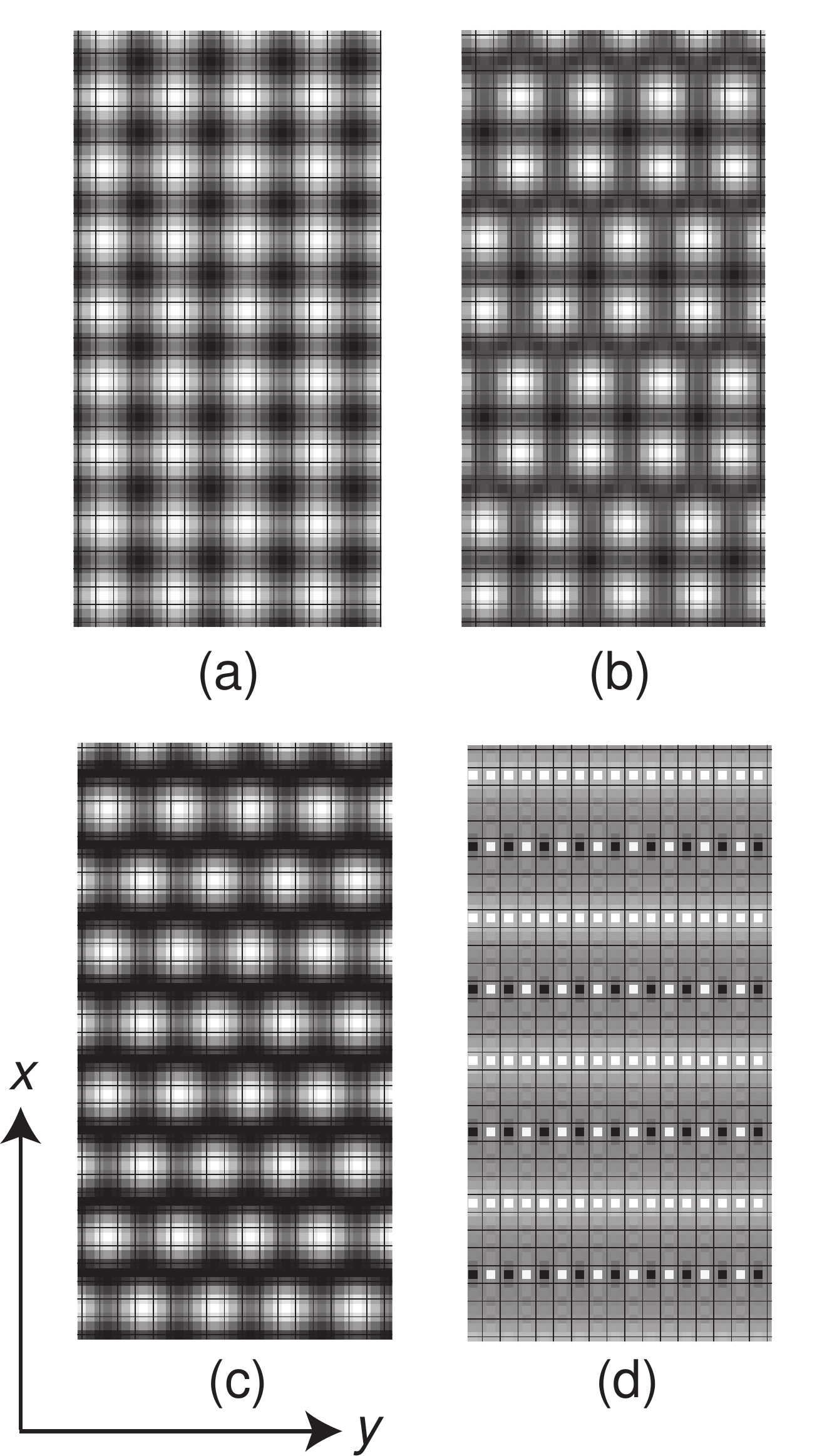}
  \caption{Structure of possible insulating states on the square lattice for paired electrons
  at a hole density of $x=1/8$ ($f=1/16$) from Ref.~\onlinecite{bbbss}. The shading is a measure of the charge density; the density is modified
  on bonds or plaquettes for configurations like those of the insulating states in Fig.~\ref{half}.}
  \label{q16}
\end{figure}

In the hole-doped cuprate superconductors, there are a variety of experimental indications \cite{jtran} 
of a stable insulator in some compounds
at a hole density of $x=1/8$. As we will discuss further in Section~\ref{sec:qc}, it is useful to consider a 
superfluid-insulator transition at $x=1/8$, and build a generalized phase diagram for the cuprates from it. The low energy
theory for such a transition is similar \cite{balents3} to that for bosons (representing paired holes) at $f=1/16$. 
We show the spatial structure of some of the large number of allowed insulating states that are possible
near the onset of superfluidity in Fig.~\ref{q16}; these are analogs of the two distinct possible state
for $f=1/2$ in Fig.~\ref{half}.
It is our proposal \cite{vojta,rst2,bbbss} that the bond-centered modulations in the local density of states
observed by Kohsaka {\em et al.\/} \cite{kohsaka} (see Fig.~\ref{phasediag} below) can be described by theories similar
to those used to obtain Fig.~\ref{q16}.

Having described the superfluid-insulator transition, we briefly turn our attention to the analogs of the `exotic' spin liquid states of magnets
discussed earlier in Sections~\ref{sec:z2} and~\ref{sec:qcp}. The topological
order associated with such states is not restricted to insulators with no broken symmetry, but can be extended not only into
insulators with conventional broken symmetries, 
but also into superfluids \cite{sf}. Indeed, an exotic metallic state is also possible \cite{ffl}, called the fractionalized Fermi liquid (FFL).
The simplest realization of a FFL appears in two band model (a Kondo lattice), in which one band forms a $Z_2$ spin liquid,
while the other band is partially filled and so forms a Fermi surface of ordinary electron-like quasiparticles. One unusual property
of a FFL is that the Fermi surface encloses a volume which does not count all the electrons: a density of exactly
one electron per unit cell is missing, having been absorbed into the $Z_2$ spin liquid. There is, thus, a close
connection between the quantum entanglement of a spin liquid and deviations from the Luttinger counting of the 
Fermi surface volume. The FFL also has interesting quantum critical points towards conventional metallic phases.
These have been reviewed in Ref.~\onlinecite{kl} and have potential applications to quantum phase transitions in the rare-earth
`heavy fermion' compounds.

Another class of metallic and superconducting states, the `algebraic charge liquids' \cite{acl} are obtained by adding mobile
carriers to antiferromagnets described by $\mathcal{S}_z$. In this case, the charge carriers are spinless fermions
which carry charges of the $A_\mu$ gauge field. These fermions can form Fermi surfaces or pair into a superconductor. 
These exotic metallic and superconducting states have a number of interesting properties which have been argued
to describe the underdoped cuprates \cite{acl}.

\section{Finite temperature quantum criticality}
\label{sec:qc}

We now turn to an important question posed in Section~\ref{sec:intro}: what is the influence of a
zero temperature quantum critical point on the $T>0$ properties ?

\begin{figure}[htbp]
  \centering
  \includegraphics[width=4.5in]{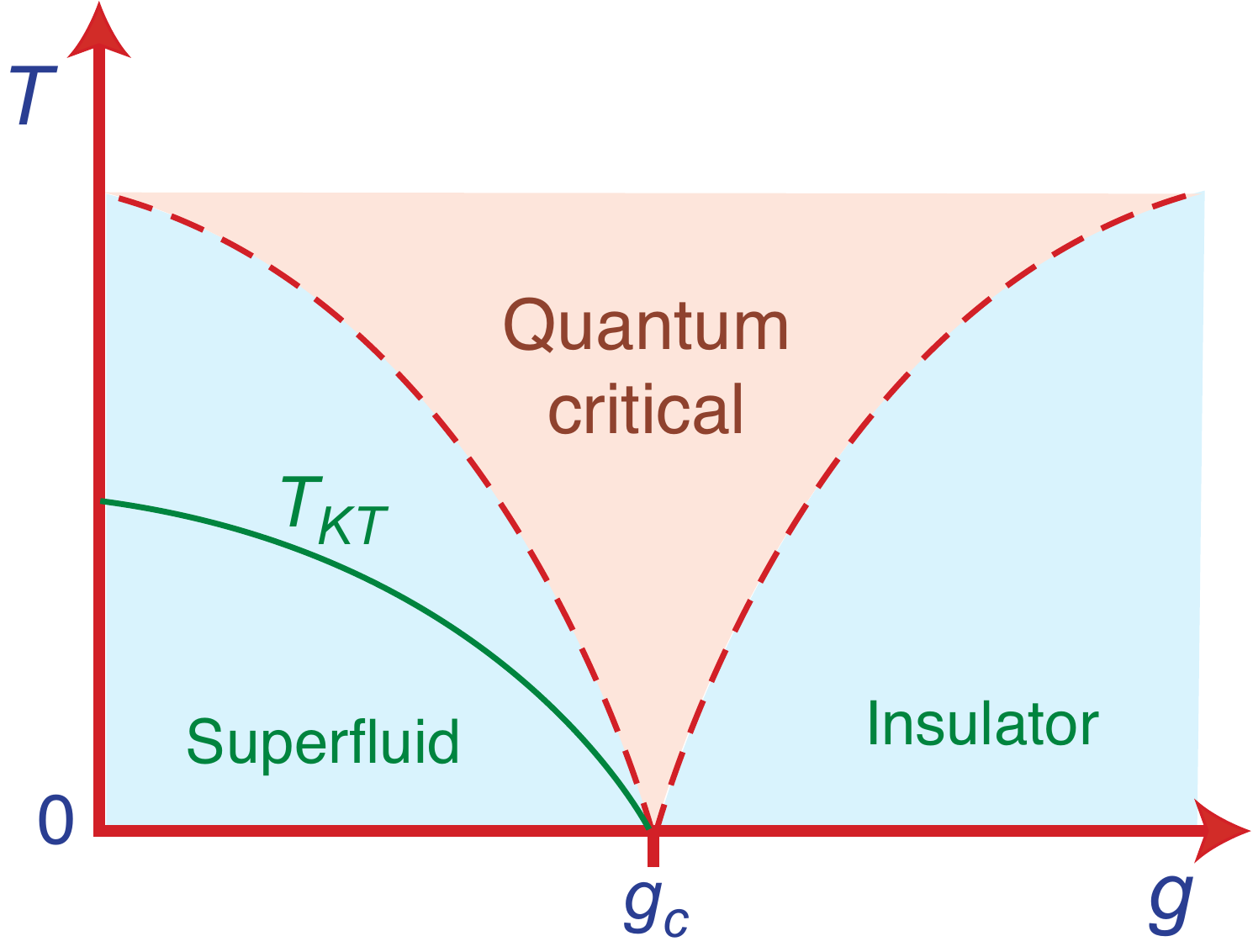}
  \caption{Phase diagram of the superfluid-insulator transition in two dimensions. Similar crossover diagrams
  apply to most of the quantum phase transitions discussed in this paper. The quantum critical point is
  at $g=g_c$, $T=0$. The dashed lines are crossovers,
  while the full line is a phase transition at the Kosterlitz-Thouless temperature $T_{KT}>0$.}
  \label{qcfig}
\end{figure}

Let us address this question in the context superfluid-insulator transition of bosons on a square lattice
at $f=1$ discussed in Section~\ref{sec:ms}. The phase diagram of this model is shown in Fig.~\ref{qcfig}
as a function of $T$ and the tuning parameter $g$ (the strength of the periodic potential). 
Note the region labeled `quantum critical': here the physics is 
described by taking the critical Hamiltonian at $g=g_c$ and placing it at a temperature $T$. The deviation of the coupling
$g$ from $g_c$ is a subdominant perturbation---upon examining the quantum fluctuations with decreasing energy scales,
the excitations of the quantum critical state are first damped out by thermal effects, before the system is
able to decide which side of the transition it resides on {\em i.e.\/}  before it is sensitive to whether $g$ is smaller
or larger than $g_c$. Thus we reach the seemingly paradoxical conclusion that the window of influence of the quantum critical
point actually {\em widens\/} as temperature is raised. Of course, there is also an upper-bound in temperature, whose value 
depends upon all details of the particular experimental system, above which the quantum criticality is irrelevant because thermal
effects are so strong that neighboring quantum degrees of freedom barely detect each other's presence.

The `transport' properties of systems in the quantum critical region are of particular interest. For the superfluid-insulator
transition, this means the study of the electrical conductivity. More generally, we can consider the response
functions associated with any globally conserved quantity, such as the total spin of the quantum magnetic systems
of Section~\ref{sec:ins}. In traditional condensed matter physics, transport is described by identifying the low-lying
excitations of the quantum ground state, and writing down `transport equations' for the conserved charges carried by them.
Often, these excitations have a particle-like nature, such as the `triplon' particles of Section~\ref{sec:dimer}, in which case
we would describe their collisions by a Boltzmann equation. In other cases, the low-lying excitations are waves, such as the spin-waves
of the N\'eel states of Section~\ref{sec:neel}, and their transport is described by a non-linear wave equation (such as the 
Gross-Pitaevski equation). 
However, for many interesting quantum critical points, including the superfluid-insulator
transition in two dimensions, neither description is possible, because the excitations do not have a particle-like
or wave-like character. 

Despite the absence of an intuitive description of the quantum critical dynamics, we can expect that the transport
properties should have a universal character determined by the quantum field theory of the quantum critical point. In addition
to describing single excitations, this field theory also determines the $S$-matrix of these excitations by the renormalization
group fixed-point value of the couplings, and these should be sufficient to determine transport properties \cite{damle}. 
So the transport co-efficients, like the conductivity, should have universal values, analogous to the universality of critical
exponents in classical critical phenomena.

Let us now focus on critical points described by CFTs, and on the transport properties of CFTs at $T>0$. CFTs in 
spatial dimension $d=1$ are especially well understood and familiar, and so we consider {\em e.g.}
the superfluid-insulator transition in one dimensional systems. A phase diagram similar to that in Fig.~\ref{qcfig}
applies: the quantum critical region of interest to us here is retained, but the critical temperature for the onset
of superfluidity is suppressed to zero. We will compute the retarded
correlation function of the charge density, $\chi (k, \omega)$, where $k$
is the wavevector, and $\omega$ is frequency; the dynamic conductivity, $\sigma(\omega)$, is related to $\chi$ by the 
Kubo formula, $\sigma (\omega) = \lim_{k \rightarrow 0} (-i \omega/k^2) \chi (k,
\omega)$.  It can be shown that {\em all\/} CFTs in $d=1$ 
have the following universal $\chi$:
\begin{equation}
\chi (k, \omega)    = \frac{4 e^2}{h} K \frac{v k^2}{(v^2 k^2 - (\omega+i \eta)^2)}~~,~~\sigma(\omega) = 
\frac{4 e^2}{h} \frac{vK}{(-i\omega)}~~;~~~\mbox{$d=1$, all $\hbar\omega/k_B T$} 
\label{d1}
\end{equation}
where $e$ and $h$ are fundamental constants which normalize the response, $K$ is a dimensionless universal number
which characterizes the CFT, $v$ is the velocity of `light' in the CFT, and $\eta$ is a positive infinitesimal. This result
holds both at $T=0$ and also at $T>0$. 
Note that this is a response characteristic of freely propagating excitations at velocity $v$. Thus, despite the complexity
of the excitations of the CFT, there is collisionless transport of charge in $d=1$ at $T>0$. Of course, in reality, there are always
corrections to the CFT Hamiltonian in any experimental system, and these will induce collisions---these issues are discussed in Ref.~\onlinecite{giam}.

It was argued in Ref.~\onlinecite{damle} that the behavior of CFTs in higher spatial dimension is qualitatively different.
Despite the absence of particle-like excitations of the critical ground state, the central characteristic of the transport
is a crossover from collisionless to collision-dominated transport. At high frequencies or low temperatures,
we can obtain a collisionless response analogous to Eq.~(\ref{d1}), using similar arguments based upon conformal invariance,
\begin{equation}
\chi (k, \omega)    = \frac{4 e^2}{h} K \frac{k^2}{\sqrt{v^2 k^2 - (\omega+i \eta)^2}}~~,~~\sigma(\omega) = 
\frac{4 e^2}{h} K ~~;~~~\mbox{$d=2$, $\hbar\omega \gg k_B T$,} 
\label{d2c}
\end{equation}
where again $K$ is a universal number.
However, it was argued \cite{damle} that phase-randomizing collisions were intrinsically present in CFTs in $d=2$
(unlike $d=1$),
and these would lead to relaxation of perturbations to local equilibrium and the consequent emergence
of hydrodynamic behavior. So at low frequencies, we have instead an Einstein relation which determines the conductivity with \cite{long-time}
\begin{equation}
\chi (k, \omega)  = 4 e^2 \chi_c \frac{Dk^2}{Dk^2 - i \omega}~~,~~\sigma(\omega) = 4 e^2 \chi_c D = 
\frac{4 e^2}{h} \Theta_1\Theta_2~~;~~~\mbox{$d=2$, $\hbar\omega \ll k_B T$,} 
\label{d2h}
\end{equation}
where $\chi_c$ is the compressibility and $D$ is the charge diffusion constant of the CFT, and these obey
$\chi_c = \Theta_1 k_B T/(h v)^2$ and $D = \Theta_2 h v^2/(k_B T)$ where $\Theta_{1,2}$ are universal numbers.
A large number of papers in the literature, 
particularly those on critical points in quantum Hall systems, have used the collisionless method of Eq.~(\ref{d2c}) to 
compute the conductivity.
However, the correct d.c. limit is given by Eq.~(\ref{d2h}); given the distinct physical interpretation of the two regimes,
we expect that $K \neq \Theta_1 \Theta_2$, and so a different conductivity in this limit.
This has been shown in a resummed perturbation expansion for a number of 
CFTs \cite{ssbook}.

Much to the surprise of condensed matter physicists \cite{jan},
remarkable progress has become possible recently on the above questions as a result of advances in string theory.  
The AdS/CFT correspondence connects the $T>0$ properties of a CFT to a dual theory of quantum gravity
of a black hole in a negatively curved ``anti-de Sitter'' universe \cite{sv,maldacena,gubser,witten,PSS}: the temperature and entropy of the CFT
determine the Hawking temperature and Bekenstein entropy of the black hole. This connection has been especially
powerful for a particular CFT, leading to the first exact results for a quantum-critical point in 2+1 dimensions.
The theory is a Yang-Mills non-Abelian gauge theory with a SU($N$) gauge group (similar to quantum chromodynamics)
and $\mathcal{N}=8$ supersymmetry. In 2+1 dimensions the Yang Mills coupling $g$ flows under
the renormalization group to a strong-coupling fixed point $g=g_c$: thus such a theory is generically quantum critical
at low energies, and in fact realizes a critical spin liquid similar to those considered in Section~\ref{sec:dcp}. 
The CFT describing this fixed point is dual to a particular phase of M theory, and this connection
allows computation, in the limit of large $N$, 
of the full $\chi (k, \omega)$ of a particular conserved density. These results contain \cite{m2}
the collisionless to hydrodynamic crossover postulated above between exactly the forms for $\chi (k, \omega)$ in Eq.~(\ref{d2c}) and (\ref{d2h}), including
the exact values \cite{CH,m2} of the numbers $K$, $\Theta_1$, $\Theta_2$. A curious feature is that
for this solvable theory, $K=\Theta_1 \Theta_2$, and this was traced to a special electromagnetic self-duality property in M theory, which
is not present in other CFTs.
It is worth noting that these are the first exact results for the conductivity of an
interacting many body system (there are also conjectured exact results for the spin diffusivity
of one-dimensional antiferromagnets with a spin gap \cite{damless}, in a phase similar to 
Section~\ref{sec:dimer}, whose dynamics was observed in Ref.~\onlinecite{collin}).

\subsection{Hydrodynamics}
\label{sec:hydro}

The transport results for the superfluid-insulator transition 
presented so far have an important limitation: they are valid only at the specific particle densities  
of the insulators, where there are a rational number of particles per unit cell 
{\em e.g.\/} the insulators in Figs.~\ref{integer}, \ref{half} and~\ref{q16}. While the particle density is 
indeed pinned at such commensurate values in the insulator at $T=0$, there is no pinning of  the density in the superfluid
state, or anywhere in the phase diagram at $T>0$. Even at small variations in density from the commensurate value
in these states, we expect strong modifications of the transport results above. This is because Eqs.~(\ref{d2c}) and (\ref{d2h})
relied crucially \cite{damle} on the lack of mixing between the conserved momentum current (in which the particle and hole excitations are
transported in the same direction) and the electrical current (which is not conserved because of collisions between particles
and holes moving in opposite directions). When particle-hole symmetry is broken by a variation in the density,
the mixing between these currents is non-zero, and a new analysis is necessary.

\begin{figure}[htbp]
  \centering
  \includegraphics[width=5.5in]{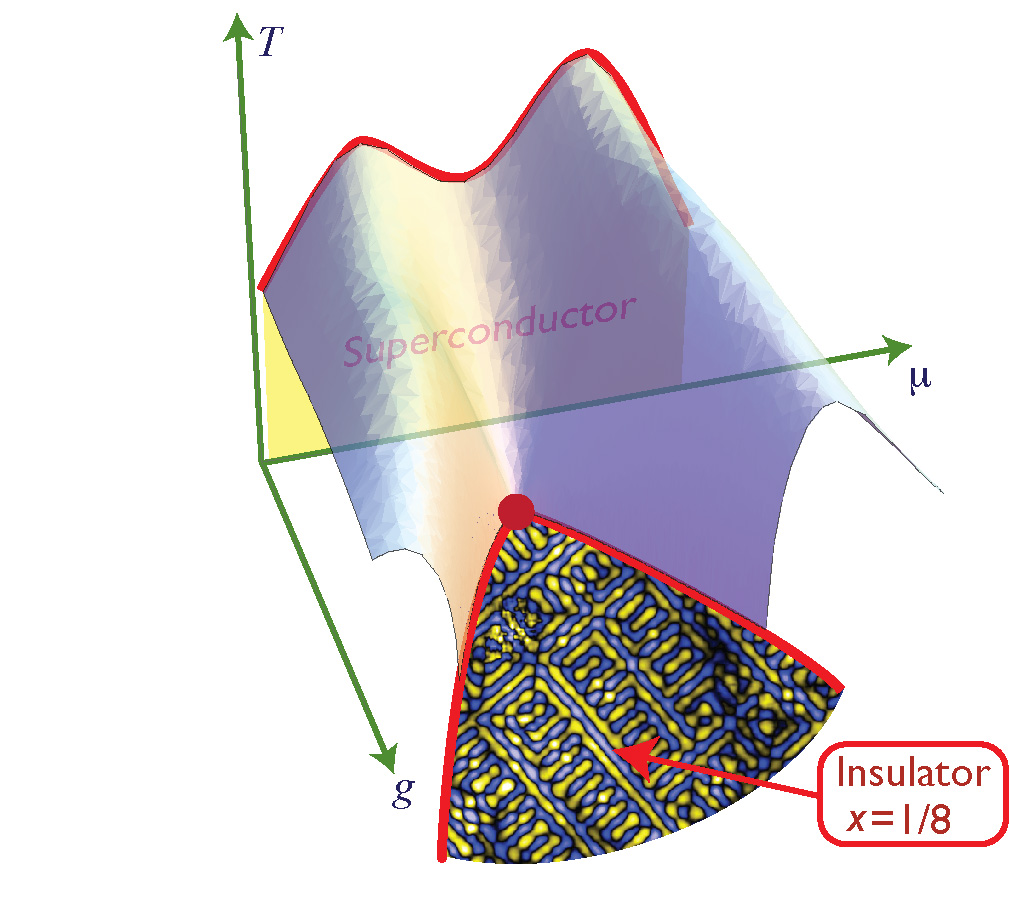}
  \caption{Superfluid-insulator transition applied to the cuprate superconductors. The parameter $g$ is as in Fig.~\ref{qcfig}, and that figure applies only exactly at hole density $x=1/8$. The above figure shows how the quantum
  critical point (red circle) controls the phase diagram at general chemical potential $\mu$ and hole density $x$. We hypothesize that
  the density modulation in the insulator is similar to the bond-centered modulations 
  in the underdoped superconductor observed by Kohsaka {\em et al.} \cite{kohsaka}, 
  as indicated by the picture of their observations. Possible supersolid
  phases are not shown.}
  \label{phasediag}
\end{figure}

As a particular motivation for such an analysis, we present in Fig.~\ref{phasediag} a hypothetical phase diagram for the
hole-doped cuprate superconductors as a function of a chemical potential, $\mu$, which controls the hole density.
We also have a tuning parameter, $g$, which controls the proximity to an insulating state whose features have presumably been
measured in Refs.~\onlinecite{kohsaka} and~\onlinecite{jtran}. We are interested in using the quantum critical 
point shown in Fig.~\ref{phasediag} to describe transport in the analog of the quantum critical region of Fig.~\ref{qcfig}
when extended away to generic densities. Further, it is also interesting to add impurities and a magnetic field, $B$, to be able to describe various magnetotransport experiments \cite{nernst}.

Further motivation for the above analysis arises from potential applications to graphene \cite{msg}. The electronic excitations of graphene
have a Dirac-like spectrum, and after including their Coulomb interactions, we obtain a system with many similarities
to quantum critical states \cite{grapheneson}. At zero bias, we expect the conductivity of pure graphene to be described
by Eqs.~(\ref{d2c}) and (\ref{d2h}) (with logarithmic corrections from the marginally irrelevant Coulomb 
interactions \cite{ssbook}), but we are interested in the bias dependence of the conductivity, and also in the
influence of impurity scattering.

A general theory of quantum critical transport with all of the above perturbations has not been achieved.
However, in the low frequency, collision-dominated regime of Eq.~(\ref{d2h}), it has recently been shown \cite{markus}
that a nearly complete description can be obtained by using general hydrodynamic arguments based upon the positivity
of entropy production and relaxation to local equilibrium. The results were verified in an explicit solution of the hydrodynamics
of a particular CFT solvable via the AdS/CFT correspondence: after the addition of a chemical potential and a magnetic field, 
the CFT maps onto a black hole with electric and magnetic charges. 

The complete hydrodynamic analysis can be found in Ref.~\onlinecite{markus}, along with a comparison
to experimental measurements of the Nernst effect in the cuprate superconductors \cite{nernst}; reasonable
agreement is obtained, with only two free parameters. The analysis is intricate, but is mainly 
a straightforward adaption of the classic procedure outlined by Kadanoff and Martin \cite{km} to the relativistic field
theories which describe quantum critical points. We list the steps: ({\em i\/}) Identify the conserved quantities, which are the energy-momentum tensor, $T^{\mu\nu}$, and the particle number current, $J^\mu$. ({\em ii\/}) Obtain the real time 
equations of motion, which express the conservation laws:
\begin{equation}
\partial_\nu T^{\mu\nu} = F^{\mu\nu}J_\nu~~~,~~~\partial_\mu J^\mu = 0;
\end{equation}
here $F^{\mu\nu}$ represents the externally applied electric and magnetic fields which can change the net
momentum or energy of the system, and we have not written a term describing momentum relaxation
by impurities. ({\em iii\/}) Identify the state variables which characterize the local thermodynamic state---we choose these to be the density, $\rho$, the temperature $T$, and an average velocity $u^\mu$. 
({\em iv\/}) Express $T^{\mu\nu}$ and $J^\mu$ in terms of the state variables and their spatial and temporal
gradients; here we use the properties of the observables under a boost by the velocity $u^\mu$, and thermodynamic quantities like the energy density, $\varepsilon$, and the pressure,
$P$, which are determined from $T$ and $\rho$ by the equation of state of the CFT. We also introduce transport 
co-efficients associated with the gradient terms. ({\em v\/}) Express the equations of motion in terms
of the state variables, and ensure that the entropy production rate is positive \cite{ll}. This is a key step which ensures
relaxation to local equilibrium, and leads to important constraints on the transport co-efficients. In $d=2$, it was found that situations with the velocity $u^\mu$ spacetime independent are characterized
by only a {\em single\/} independent transport co-efficient \cite{markus}. This we choose to be the longitudinal conductivity
at $B=0$.  ({\em vi\/}) Solve the initial value problem for the state variables using the linearized equations
of motion. ({\em vii\/}) Finally, translate this solution to the linear response functions, as described in Ref.~\onlinecite{km}.

As an example of the above analysis, we
highlight an important result--the modification
of the conductivity in Eq.~(\ref{d2h}) by a non-zero density difference, $\rho$, from that of the Mott insulator at $B=0$.
We found that a non-zero $\rho$ introduces an additional component of the current which cannot relax by
collisions among the excitations, and which requires the presence of impurities to yield a finite conductivity:
\begin{equation}
\sigma (\omega) = \frac{4 e^2}{h} \Theta_1 \Theta_2 + 
\frac{4 e^2 \rho^2 v^2}{(\varepsilon+P)} \frac{1}{(-i \omega + 1/\tau_{\rm imp})}.
\label{sigmatau}
\end{equation}
Here the first term is as in Eq.~(\ref{d2h}), arising from the collisions between the particle and hole
excitations of the underlying CFT, although the values of $\Theta_1$ and $\Theta_2$ can change as function
of $\rho$. The additive
Drude-like contribution arises from the excess particles or holes induced by the non-zero $\rho$,
and $1/\tau_{\rm imp}$
is the impurity scattering rate (which can also be computed from the CFT \cite{nernst,impure}). 
Note that in the `non-relativistic limit', achieved when $\rho$ is so large
that we are well away from the critical point, the energy and pressure are dominated by the rest
mass of the particles $\varepsilon+P \approx |\rho| m v^2$, and then the second term reduces
to the conventional Drude form. The above theory can also be extended to include the Hall response
in a magnetic field, $B$, \cite{sean1,markus}; in particular we obtain the d.c. Hall conductivity, $\sigma_{xy}
\propto \rho/B$. This predicts a negative Hall resistance in the cuprates at hole densities smaller than that of the 
insulator at $x=1/8$, as has been observed in recent experiments on the cuprates \cite{louis}.

\section{Conclusions}

We have described progress in understanding new types of entanglement in many electron systems.
Entanglement is a key ingredient for quantum computing, and the critical spin liquids described here are likely
the states with most complex entanglement known.
This progress has been made possible by advances on several distinct fronts. On the experimental side,
new `correlated electron' materials have been discovered, such as those described in Refs.~\onlinecite{kato1,shlee}.
There have been advances in experimental techniques, such as scanning tunneling microscopy in Ref.~\onlinecite{kohsaka},
and neutron scattering in Refs.~\onlinecite{collin}. And a new frontier has opened in the study of correlated phase
of cold atoms in optical lattices \cite{bloch}. On the theoretical side, large scale numerical studies of new types of quantum
magnets has become possible by new algorithms, including the first studies of phases with no magnetic order in $S=1/2$ 
square lattice antiferromagnets \cite{anders,rkk,shailesh}. An important frontier is the extension of such methods
to a wider class of frustrated quantum systems, which have so far proved numerically intractable because of the `sign'
problem of quantum Monte Carlo. In field-theoretical studies of novel quantum phases, new connections have emerged
to those studied in particle physics and string theory\cite{jan,m2,markus}, and this promises an exciting cross-fertilization between the two fields. 
\acknowledgments

I thank Sander Bais, Sean Hartnoll, Ribhu Kaul, Pavel Kovtun, Roger Melko, Max Metlitski, Markus M\"uller, and Anders Sandvik for valuable comments. 
This work was supported by NSF Grant No.\ DMR-0537077.


\begin{thebibliography}{99}

\bibitem{epr} Einstein,~A., Podolsky,~B. \& Rosen,~N.
Can quantum mechanical description of physical reality be considered
  complete? {\em Phys.\ Rev.\/}  {\bf 47}, 777-780 (1935).

\bibitem{matsumoto} Matsumoto,~M., Yasuda,~C., Todo,~S. \&
Takayama,~H. Ground-state phase diagram of quantum Heisenberg
antiferromagnets on the anisotropic dimerized square lattice.
{\em Phys. Rev. B\/} {\bf 65}, 014407 (2002).

\bibitem{chn} Chakravarty,~A. Halperin,~B.~I. \& Nelson,~D.~R. 
Two-dimensional quantum Heisenberg antiferromagnet at low temperatures.
{\em Phys. Rev. B\/} {\bf 39}, 2344-2371 (1989). 

\bibitem{csy}  Chubukov,~A.~V., Sachdev,~S. \& Ye,~J. 
Theory of two-dimensional quantum antiferromagnets with a nearly-critical ground state. {\em Phys. Rev. B} {\bf 49}, 11919-11961 (1994). 

\bibitem{coldea} Coldea,~R., Tennant,~D.~A. \& Tylczynski,~Z. 
Extended scattering continua characteristic of spin fractionalization in the two-dimensional frustrated quantum 
magnet Cs$_2$CuCl$_4$ observed by neutron scattering. {\it Phys. Rev. B\/} {\bf 68}, 134424 (2003). 

\bibitem{bernu} Bernu,~B., Lecheminant,~P., Lhuillier,~C. \& Pierre,~L. 
Exact spectra, spin susceptibilities, and order parameter of the quantum Heisenberg antiferromagnet on the triangular lattice. {\it Phys. Rev. B} {\bf 50}, 10048-10062 (1994).

\bibitem{css} Chubukov,~A.~V., Senthil,~T. \& Sachdev,~S.
Universal magnetic properties of frustrated quantum antiferromagnets in two dimensions.
{\em Phys. Rev. Lett.\/} {\bf 72}, 2089-2092 (1994).

\bibitem{oosawa} Oosawa,~A., Fujisawa,~M., Osakabe,~T., Kakurai,~K. \& Tanaka,~H. Neutron Diffraction Study of the Pressure-Induced Magnetic
Ordering in the Spin Gap System TlCuCl$_3$. {\it J. Phys. Soc. Jpn\/}
{\bf 72}, 1026-1029 (2003).

\bibitem{ruegg} R\"uegg,~Ch., Cavadini,~N., Furrer,~A., G\"udel,~H.-U., Kr\"amer,~K., Mutka,~H., 
Wildes,~A., Habicht,~K. \& Vorderwisch,~P.
Bose-Einstein condensation of the triplet states in the magnetic
insulator TlCuCl$_3$. {\it Nature\/} {\bf 423}, 62-65 (2003).

\bibitem{jaime} Jaime,~M. {\em et al.\/} Magnetic-Field-Induced Condensation of Triplons 
in Han Purple Pigment BaCuSi$_2$O$_6$. {\it Phys. Rev. Lett.\/} {\bf 93}, 087203 (2004). 

\bibitem{cavadini} Cavadini,~N. {\em et al.} Magnetic excitations in the quantum spin system TlCuCl$_3$.
{\it Phys. Rev. B} {\bf 63}, 172414 (2001).

\bibitem{collin} Xu,~G., Broholm,~C., ~Soh,~Y.-A., Aeppli,~G., DiTusa,~J.~F., Chen.~Y., Kenzelmann,~M., Frost,~C.~D., Ito,~T., Oka,~K. \& Takagi,~H. Mesoscopic Phase Coherence in a Quantum Spin Fluid.
{\it Science\/} {\bf 317}, 1049-1052 (2007).


\bibitem{kato1} Tamura,~M., Nakao,~A. \& Kato,~R. Frustration-Induced Valence-Bond Ordering
in a New Quantum Triangular Antiferromagnet
Based on [Pd(dmit)$_2$]. {\it J. Phys. Soc. Japan\/} {\bf 75}, 093701 (2006).

\bibitem{shlee} Lee,~S.-H., Kikuchi,~H. Qiu,~Y., Lake,~B., Huang,~Q., Habicht,~K. \& Kiefer,~K.
Quantum-spin-liquid states in the two-dimensional kagome antiferromagnets Zn$_x$Cu$_{4-x}$(OD)$_6$Cl$_2$. {\it Nature Materials\/} {\bf 6}, 853-857 (2007).

\bibitem{kohsaka} Kohsaka,~Y., Taylor,~C., Fujita,~K., Schmidt,~A., Lupien,~C., 
Hanaguri,~T., Azuma,~M., Takano,`M., Eisaki,~H., Takagi,~H., Uchida,~S. \& Davis,~J.C. 
An Intrinsic Bond-Centered Electronic Glass with Unidirectional Domains in Underdoped Cuprates.
{\it Science\/} {\bf 315}, 1380-1385 (2007).

\bibitem{poilblanc} Raczkowski,~M., Capello,~M., Poilblanc,~D., Fr\'esard,~R. \& Ol\'es,~A.~M.  
Unidirectional $d$-wave superconducting domains in the two-dimensional $t$-$J$ model.
{\it Phys. Rev. B\/} {\bf 76}, 140505(R) (2007).

\bibitem{vojta}  
Vojta,~M. \& R\"osch,~O. 
Superconducting $d$-wave stripes in cuprates: Valence bond order coexisting with nodal quasiparticles.
arXiv:0709.4244.

\bibitem{rst2} Sachdev,~S. \& Read,~N. Large $N$ expansion for frustrated and doped quantum 
antiferromagnets. {\em Int. J. Mod. Phys. B\/} {\bf 5}, 219Ð249 (1991).

\bibitem{anders}  Sandvik,~A.~W., {\em Evidence for Deconfined Quantum Criticality in a Two-Dimensional Heisenberg Model with Four-Spin Interactions},
Phys. Rev. Lett. {\bf 98}, 227202 (2007).

\bibitem{rkk} Melko,~R.~G. \& Kaul,~R.~K., Universal Scaling in the Fan of an Unconventional Quantum Critical Point.
{\it Phys. Rev. Lett.\/} 100, 017203 (2008).

\bibitem{shailesh} Jiang,~F.-J., Nyfeler,~M., Chandrasekharan,~S. \& Wiese,~U.-J.
From an Antiferromagnet to a Valence Bond Solid: Evidence for a First Order Phase Transition. 
arXiv:0710.3926.

\bibitem{roger} Melko,~R.~G., Sandvik,~A.~W. \& Scalapino,~D.~J.
Two-dimensional quantum XY model with ring exchange and external field.
{\it Phys. Rev. B\/} {\bf 69}, 100408(R) (2004).

\bibitem{rsl} Read,~N. \& Sachdev,~S. Valence bond and spin-Peierls
ground states of low-dimensional quantum
antiferromagnets. {\em Phys. Rev. Lett.\/} {\bf 62}, 1694-1697 (1989); {\em ibid.\/}
Spin-Peierls, valence-bond solid, and N\'eel ground-states of low-dimensional quantum antiferromagnets.
{\em Phys. Rev. B\/} {\bf 42}, 4568-4569 (1990).

\bibitem{mv}  Motrunich,~O.~I. \& Vishwanath,~A. 
Emergent photons and transitions in the O(3) sigma model with hedgehog suppression.
{\it Phys. Rev. B\/} {\bf 70}, 075104 (2004).

\bibitem{fk} Fradkin,~E. \& Kivelson,~S. Short range resonating valence bond theories and
superconductivity. {\it Mod. Phys. Lett. B\/} {\bf 4}, 225-232 (1990).

\bibitem{polyakov} Polyakov,~A.~M. {\em Gauge fields and strings\/} (Harwood Academic, New York, 1987).

\bibitem{haldane} Haldane,~F.~D.~M. O(3) Nonlinear $\sigma$ Model and
the Topological Distinction between Integer- and Half-Integer-Spin
Antiferromagnets in Two Dimensions. {\it Phys. Rev. Lett.\/} {\bf 61},
1029-1032 (1988).

\bibitem{senthil2} Senthil,~T., Balents,~L.,
Sachdev,~S., Vishwanath,~A. \& Fisher,~M.~P.~A. Quantum
criticality beyond the Landau-Ginzburg-Wilson paradigm. {\em Phys.
Rev. B\/} {\bf 70}, 144407 (2004).

\bibitem{rst1} Read,~N. \& Sachdev,~S. Large $N$ expansion for frustrated quantum antiferromagnets. {\em Phys.  
Rev. Lett.\/} {\bf 66}, 1773Ð1776 (1991).

\bibitem{sstri} Sachdev,~S., 
Kagome and triangular lattice Heisenberg antiferromagnets: ordering from quantum fluctuations and quantum-disordered 
ground states with deconfined bosonic spinons.
{\em Phys. Rev. B\/} {\bf 45}, 12377-12396 (1992).

\bibitem{wen} Wen,~X.-G. Mean-field theory of spin-liquid states with finite energy gap and topological orders.
{\em Phys. Rev. B\/} {\bf 44}, 2664-2672 (1991).

\bibitem{jalabert} Jalabert,~R. \& Sachdev,~S. Spontaneous alignment of frustrated bonds in an anisotropic, three dimensional Ising model. {\em Phys. Rev. B\/} {\bf 44}, 686-690 (1991). 

\bibitem{vs} Sachdev,~S. \& Vojta,~M. Translational symmetry breaking in two-dimensional antiferromagnets and superconductors. J. Phys. Soc. Japan {\bf 69}, Suppl. B, 1-9 (2000).

\bibitem{bais1} Bais,~F.~A. Flux metamorphosis. {\em Nucl. Phys. B\/} {\bf 170}, 32Ð43 (1980).

\bibitem{bais2} Bais,~F.~A., van Driel,~P. \& de Wild Propitius, M. Quantum symmetries in discrete gauge 
theories. {\em Phys. Lett. B\/} 280, 63Ð70 (1992).

\bibitem{kitaev} Kitaev,~A.~Y.
Fault-tolerant quantum computation by anyons.
{\em Annals of Physics\/}  {\bf 303}, 2-30 (2003).

\bibitem{preskill}
  Wang, C., Harrington, J. \& Preskill, J.
  Confinement-Higgs transition in a disordered gauge theory and the accuracy
  threshold for quantum memory.
  {\em Annals of Physics \/}  {\bf 303}, 31-58 (2003).
  
\bibitem{wen2} Wen,~X.-G. Quantum orders in an exact soluble model.
{\em Phys. Rev. Lett.\/} {\bf 90}, 016803 (2003).

\bibitem{sf} Senthil,~T. \& Fisher,~M.~P.~A. $Z_2$ gauge theory of electron fractionalization in strongly correlated systems. {\em Phys. Rev. B} {\bf 62}, 7850-7881 (2000).

\bibitem{sondhi} Moessner,~R. \& Sondhi,~S.~L. 
Resonating Valence Bond Phase in the Triangular Lattice Quantum Dimer Model.
{\it Phys. Rev. Lett.\/} {\bf 86}, 1881-1884 (2001).

\bibitem{freedman} Freedman,~M., Nayak,~C., Shtengel,~K., Walker,~K. \& Wang,~Z.
A class of P,T-invariant topological phases of interacting electrons.
{\it Annals of Physics\/} {\bf 310}, 428-492 (2004).


\bibitem{ruegg2} R\"uegg,~Ch., Normand,~B., Matsumoto,~M., 
Furrer,~A., McMorrow,~D., Kr\"amer,~K., G\"udel,~H.-U., 
Gvasaliya,~S., Mutka,~H. \& Boehm,~M. 
Pressure-Controlled Quantum Fluctuations and
Elementary Excitations in Quantum Magnets. preprint.

\bibitem{fisherliu} Liu,~K.-S. \& Fisher,~M.~E. Quantum lattice gas and
the existence of a supersolid. {\it J. Low. Temp. Phys.\/} {\bf 10}, 655-683 
(1973).

\bibitem{senthil1} Senthil,~T., Vishwanath,~A., Balents,~L.,
Sachdev,~S. \& Fisher,~M.~P.~A. Deconfined Quantum Critical
Points. {\em Science\/} {\bf 303}, 1490-1494 (2004).


\bibitem{kuklov} Kuklov,~A.~B., ProkofÕev,~N.~V., Svistunov,~B.~V. \& Troyer,~M. 
Deconfined criticality, runaway flow in the 
two-component scalar electrodynamics and weak first-order superfluid-solid transitions.
{\it Annals of Physics\/} {\bf 321}, 1602-1621 (2006).


\bibitem{flavio} Nogueira,~F.~S., Kragset,~S. \& Sudbo,~A.
Quantum critical scaling behavior of deconfined spinons. {\em Phys. Rev. B\/} {\bf 76}, 220403(R) (2007).

\bibitem{rantwen} Rantner,~W. \& Wen,~X.-G.
Electron spectral function and algebraic spin liquid for the normal state of underdoped high $T_c$ superconductors.
{\it Phys. Rev. Lett.\/} {\bf 86}, 3871-3874 (2001).

\bibitem{brad} Affleck,~I. \& Marston,~J.~B. 
Large-$n$ limit of the Heisenberg-Hubbard model: Implications for high-$T_c$ superconductors.
{\it Phys. Rev. B \/} {\bf 37}, 3774-3777 (1988).

\bibitem{stableu1} Hermele,~M., Senthil,~T., Fisher,~M.~P.~A., Lee,~P.~A., Nagaosa,~N. \& Wen,~X.-G.
Stability of {\rm U(1)} spin liquids in two dimensions.
{\it Phys. Rev. B\/} {\bf 70}, 214437 (2004).

\bibitem{motherasl} Hermele,~M., Senthil,~T. \& Fisher,~M.~P.~A.
Algebraic spin liquid as the mother of many competing orders.
{\it Phys. Rev. B\/} {\bf 72} 104404 (2005).

\bibitem{aslkagome} Ran,~Y., Hermele,~M., Lee,~P.~A. \& Wen,~X.-G.
Projected wavefunction study of Spin-1/2 Heisenberg model on the Kagome lattice.
{\it Phys. Rev. Lett.\/} {\bf 98}, 117205 (2007).

\bibitem{bloch} Bloch.~I.Ultracold quantum gases in optical lattices.
{\it Nature Physics\/} {\bf 1}, 23-30 (2005).

\bibitem{fwgf} Fisher,~M.~P.~A., Weichmann,~P.~B., Grinstein,~G. \& Fisher,~D.~S. Boson localization and the superfluid-insulator transition.
{\it Phys.~Rev.~B\/} {\bf 40}, 546-570 (1989).

\bibitem{bbbss} Balents,~L., Bartosch,~L., Burkov,~A., Sachdev,~S. \& Sengupta,~K.
Putting competing orders in their place near the Mott
transition. {\it Phys. Rev. B\/} {\bf 71}, 144508 (2005).

\bibitem{lfs} Lannert,~L., Fisher,~M.~P.~A. \& Senthil,~T. Quantum confinement transition in a $d$-wave superconductor.
{\it Phys. Rev. B\/} {\bf 63}, 134510 (2001).

\bibitem{balents3} Balents,~L. \& Sachdev,~S. Dual vortex theory of doped antiferromagnets.
{\it Annals of Physics\/} {\bf 322}, 2635-2664 (2007).

\bibitem{jtran} Tranquada,~J.~M., Woo,~H., Perring,~T.~G., Goka,~H., 
Gu,~G.~D., Xu,~G., Fujita,~M. \& Yamada,~K. 
Quantum magnetic excitations from stripes in copper oxide superconductors.
{\em Nature\/} {\bf 429}, 534-538 (2004). 

\bibitem{ffl} Senthil,~T., Sachdev,~S. \& Vojta,~M. Fractionalized Fermi liquids. {\em Phys.~Rev.~Lett.\/} {\bf 90}, 216403 (2003).

\bibitem{kl} Senthil,~T., Sachdev,~S. \& Vojta,~M. Quantum phase transitions
out of the heavy Fermi liquid. {\it Physica B\/} {\bf 359-361}, 9-16 (2005).

\bibitem{acl} Kaul,~R.~K., Kim,~Y.~B., Sachdev,~S. \& Senthil,~T. Algebraic charge liquids. {\it Nature Physics\/}, {\bf 4}, 28-31 (2008).

\bibitem{damle} Damle,~K. \& Sachdev,~S. Non-zero temperature transport near
quantum critical points. {\it Phys.\ Rev.\ B\/} {\bf 56}, 8714-8733 (1997).

\bibitem{giam} Giamarchi,~T. Umklapp process and resistivity in one-dimensional fermion systems. {\it Phys. Rev. B\/} {\bf 44}, 2905-2913 (1991).

\bibitem{long-time} For the case of {\em neutral\/} boson superfluids (but not charged systems like the cuprate 
superconductors of Fig~\ref{phasediag}),
hydrodynamic `long-time tails' cause the constants $D$ and $\Theta_2$ to acquire
a weak logarithmic dependence on $\hbar \omega/k_B T$ at small $\omega$ in a sample with perfect 
momentum conservation. Kovtun,~P. \& Yaffe,~L.~G. Hydrodynamic Fluctuations, 
Long-time Tails, and Supersymmetry. {\it Phys. Rev. D\/} {\bf 68}, 025007 (2003).
 
\bibitem{ssbook} Sachdev,~S., {\em Quantum Phase Transitions}
(Cambridge University Press, Cambridge, 1999).

\bibitem{jan} Zaanen,~J. A black hole full of answers.
{\it Nature\/} {\bf 448}, 1000-1001 (2007).

\bibitem{sv}
  Strominger,~A. \& Vafa,~C.
  Microscopic Origin of the Bekenstein-Hawking Entropy.
  {\it Phys.\ Lett.\  B\/} {\bf 379}, 99-104 (1996).
  
\bibitem{maldacena} 
  Maldacena,~J.~M.
  The large $N$ limit of superconformal field theories and supergravity.
  {\it Adv.\ Theor.\ Math.\ Phys.\/}  {\bf 2}, 231-252 (1998).
  
\bibitem{gubser} Gubser,~S.~S., Klebanov,~I.~R. \& Polyakov,~A.~M.
Gauge theory correlators from non-critical string theory.
{\it Phys.\ Lett.\  B\/} {\bf 428}, 105-114 (1998).


\bibitem{witten} Witten,~E.
  Anti-de Sitter space and holography.
  {\it Adv.\ Theor.\ Math.\ Phys.\/}  {\bf 2}, 253-290 (1998).

\bibitem{PSS} Policastro,~G., Son,~D.~T. \& Starinets,~A.~O.
From AdS/CFT correspondence to hydrodynamics.
{\it JHEP\/} {\bf 0209}, 043 (2002).

\bibitem{m2}
  Herzog,~C.~P., Kovtun,~P.~K., Sachdev,~S. \& Son,~D.~T.
  Quantum critical transport, duality, and M-theory.
  {\it Phys.\ Rev.\  D\/} {\bf 75}, 085020 (2007).

\bibitem{CH}
  Herzog,~C.~P.
   The hydrodynamics of M-theory.
  {\it JHEP\/} {\bf 0212}, 026 (2002).
  
\bibitem{damless} Damle,~K. \& Sachdev,~S. Spin dynamics and transport in gapped one-dimensional Heisenberg antiferromagnets at nonzero temperatures. {\it Phys. Rev. B\/} {\bf 57}, 8307-8339 (1998). 


\bibitem{nernst}  Wang,~Y., Li,~L. \&
Ong,~N.~P. Nernst effect in high-$T_c$ superconductors. {\it Phys. Rev. B\/} {\bf 73}, 024510 (2006).

\bibitem{msg} M\"uller,~M. \& Sachdev,~S. Collective cyclotron motion of the relativistic plasma in graphene.  arXiv:0801.2970.

\bibitem{grapheneson} Son,~D.~T.
Quantum critical point in graphene approached in the limit of infinitely strong Coulomb interaction.
{\em Phys. Rev. B\/} {\bf 75}, 235423 (2007).

\bibitem{markus} Hartnoll,~S.~A., Kovtun,~P.~K., M\"uller,~M. \& Sachdev,~S. 
Theory of the Nernst effect near quantum phase transitions in condensed matter, and in dyonic black holes. {\it Phys. Rev. B\/} 
{\bf 76}, 144502 (2007).

\bibitem{km} Kadanoff,~L.~P. \& Martin,~P.~C. Hydrodynamic equations and correlation functions.
{\em Annals of
Physics\/} {\bf 24}, 419-469 (1963).

\bibitem{ll} Landau,~L.~D. \& Lifshitz,~E.~M. {\em Fluid Mechanics\/}, 
Section 127 (Butterworth-Heinemann, Oxford, 1987).

\bibitem{impure} Hartnoll,~S.~A. \& Herzog,~C.~P. Impure AdS/CFT. 
arXiv:0801.1693.

\bibitem{sean1} Hartnoll,~S.~A. \& Kovtun~P.~K.
Hall conductivity from dyonic black holes.
{\em Phys.\ Rev.\  D} {\bf 76}, 066001 (2007).


\bibitem{louis} Doiron-Leyraud, N.{\em et al.}
Quantum oscillations and the Fermi surface in an underdoped high-$T_c$ superconductor, {\em Nature\/} {\bf 447}, 565-568 (2007).

\end{thebibliography}
\end{document}